# Accurate Angle-Resolved Raman Spectroscopy Methodology – Quantifying the Dichroic Edge Filter Effect

Tehseen Adel[1,2†], Maria F. Munoz[1†], Thuc T. Mai[1], Charlezetta E. Wilson-Stokes[1,3], Riccardo Torsi[1], Kevin M. Linne von Berg[4], Aurélien Thieffry[5], Jeffrey R. Simpson[1,4], and Angela R. Hight Walker[1*]

[1]Quantum Measurement Division, National Institute of Standards and Technology, Gaithersburg, MD, USA
[2]Department of Physical Sciences, University of Findlay, Findlay, OH, USA
[3]Department of Mechanical Engineering, Howard University, Washington, DC, USA
[4]Department of Physics, Astronomy and Geosciences, Towson University, Towson, MD, USA
[5] HORIBA France, Lille, Hauts-de-France, France

**Abstract**

Angle-resolved Raman spectroscopy (ARRS) is an effective method to analyze the symmetry of phonons and other excitations in molecules and solid-state crystals. While there are several configurations of ARRS instruments, the measurement system detailed here utilizes two pairs of linear polarizers and superachromatic half-wave plates. After the orientations of the linear polarizers are set to fixed angles, the two half-wave plates rotate independently, through motorized control, enabling 2D linear polarization mapping. Described within is a protocol to achieve high quality ARRS measurements leveraging phonons from easily accessible test materials [molybdenum disulfide ($MoS_2$), sapphire ($Al_2O_3$), and silicon] to validate the system and operation. Quantitative polarized Raman data strongly depends on the quality of the sample surface and the optics: the order of placement, alignment, and any distortion caused by their coatings. This study identifies the impact of commonly used edge filters on the polarization response of materials with an anisotropic response, as emulated by the $T_{2g}$ phonon in Si(100).

* Email: angela.hightwalker@nist.gov

We detect and model the significant distortion of the $T_{2g}$ phonon polarization response originating from our dichroic edge filters, the results of which are broadly applicable to optics in any Raman instrument. This ARRS setup also enables helicity-resolved Raman measurements by replacing the first half-wave plate with a superachromatic quarter-wave plate; this configuration is also validated using the Raman response of the aforementioned test materials. This paper aims to enhance the quality and reproducibility of polarized Raman measurements through both instrumental considerations and methodological improvements.

**Keywords**

polarization, Raman spectroscopy, optics, methodology, calibration

## 1. Introduction

Angle-resolved, polarized Raman spectroscopy (ARRS) is a powerful technique for determining the symmetry of phonons in crystals and molecules. In such measurements, the polarization directions of both the incident and scattered light are varied in relation to the crystallographic axes or molecular orientation of the sample under examination. Thorough analysis of Raman spectra collected in different polarization combinations enables the assignment of mode symmetry and the determination of Raman tensor elements.[1] ARRS has not only been used for mode assignment but has also been extensively applied as a rapid, non-destructive technique to assess the orientation of anisotropic nanomaterials and correlate it with their direction-dependent properties.[2] For example, Wood *et al*. and others utilized ARRS to successfully characterize the molecular orientation in organic semiconductors and linked this to charge transport.[3,4] Similarly, He *et al.* introduced a theoretical framework involving ARRS to investigate stress and crystallographic orientation in silicon, a material extensively used in semiconductor applications.[5] Researchers also employed polarized Raman spectroscopy as a quantitative method for determining carbon nanotube orientation.[6,7] In the field of 2D materials, ARRS has become a ubiquitous method for documenting the crystal orientation of anisotropic samples like black phosphorus (BP),[8] $ReS_2$,[9,10] and $WTe_2$.[11] This capability is increasingly crucial for the development of van der Waals[12] and moiré heterostructures,[13] whose properties are highly sensitive to the relative orientation of each layer within the stack. Additionally, Yu *et al.* provided proof of concept for using polarized Raman spectroscopy measurements to develop models for quantitative defect analysis in anisotropic 2D materials.[14]

However, the practical application of ARRS is fraught with challenges that can impede its accuracy and reliability, thereby limiting its reproducibility from one laboratory to another. One significant issue is the inconsistent measurement configurations found across the literature, particularly regarding the placement of polarization optics along the laser excitation and Raman scattered paths.[15] Moreover, the impact of polarization distortion from optical components is rarely addressed, with limited discussions focusing primarily on the objective lens[16],[17] and grating[18,19]. Although the impacts of the objective lens and grating are generally well understood and accounted for in most experimental setups, such as installing a polarization scrambler in front of the grating, the effects of the dichroic edge filter have largely been overlooked. The dichroic edge filter plays a crucial role in most modern Raman spectroscopy instruments, reflecting the incident laser light towards the sample while allowing only the scattered Raman signal to pass through to the detector. The polarization aberration effects of dichroic mirrors, which arise from the difference in phase shift they impart on the $s$- and $p$-components of the incident polarization, have been investigated in other optical and microscopy applications.[20],[21] However, their effects on polarized Raman measurements have been neglected, potentially causing incorrect interpretations of polarized Raman data. Therefore, there is a pressing need to develop a methodology and experimental configuration for performing polarization measurements to ensure accurate and consistent results.

In this paper, we present a methodology for performing accurate, ARRS measurements with specific, easily accessible test materials. The three materials employed are 2H-phase molybdenum disulfide ($MoS_2$), $c$-plane sapphire ($Al_2O_3$ 001), and silicon wafers of (100), (110), and (111) cuts. $MoS_2$ and sapphire have in-plane phonons with an isotropic Raman response, specifically the $A_g$ and $E_g$ phonons. The anisotropic response of materials is represented through the (100)-oriented Si via its $T_{2g}$ phonon. We first introduce the ARRS experimental setup and systematically characterize the optical path using a polarimeter. We then explore the challenges in making accurate measurements with bulk 2H-$MoS_2$ as our calibration material, which include ensuring a flat, clean surface and accounting for any resonance effects. Our ARRS measurements on the $T_{2g}$ phonon response from Si(100) reveal that the dichroic edge filter alters the scattered polarization at certain angles, leading to discrepancies between the experimental results and theoretical simulations with an ideal instrument response. Considering this discrepancy, we demonstrate that modeling the edge filter as a waveplate allows us to accurately reproduce the experimental response. Finally, we touch on how substituting our incident half-wave plate with a quarter-wave plate enables helicity-resolved Raman spectroscopy (HRRS) experiments. This paper presents a straightforward methodology and a comprehensive investigation into the impact of the dichroic edge filter on ARRS data.

## 2. Materials and Methods

### 2.1 Materials

We use three well-known and easily accessible materials to characterize the ARRS experimental system: 2H-phase molybdenum disulfide ($MoS_2$) (99.999%, CVT grown, HQ Graphene), c-cut sapphire crystal ($Al_2O_3$(001); one-sided polished, MTI Corporation), and Si wafers of (100), (110), and (111) cuts (polished, N-type, Resistivity >1000 Ω·cm, MTI Corporation). For the Si wafers, the primary flat (longest edge on the wafer) is perpendicular to the incident light polarization, such that the angle between the incident light and the plane (100) is $45^\circ$ as shown in **Figure S1**. We selected 2H-$MoS_2$ as the primary polarization calibration material due to its strong, polarization sensitive $A_{1g}$ phonon mode (409.27 $cm^{-1}$). In our experiments, we observed some minor disadvantages to using $MoS_2$, particularly as the stability of the Raman active modes are susceptible to environmental degradation,[22] thickness dependence,[23–25] and exhibit resonant characteristics with specific excitation wavelengths.[25,26] Since $MoS_2$ is a van der Waals material, the surface of bulk $MoS_2$ can easily be "refreshed" by physically exfoliating the top layers with sticky tape. For consistent, reproducible results over extended periods, we recommend utilizing a freshly exfoliated, bulk crystal rather than a few-layer flake, because the Raman response from thin flakes is known to be affected by the substrate.[27] Using a bulk crystal will ensure reliable calibration and data collection. As an alternative calibration material, c-cut sapphire crystal also has a distinctive, polarization-active $A_{1g}$ mode (417.32 $cm^{-1}$) and requires less preparation compared to $MoS_2$. However, the intensity of its Raman response is substantially lower, requiring longer acquisition times and higher laser power exposure to the sample at any given wavelength.

Finally, we use crystalline Si wafers, particularly the $T_{2g}$ phonon (520.89 $cm^{-1}$), which is crucial for calibrating the Raman shift (wavenumbers or relative wavelength frequency on the x-axis) in Raman systems. The response of Si varies depending on the crystal cut, with Si(111) exhibiting an isotropic Raman response to incident light polarization.[28] Conversely, Si(100) and Si(110) display polarization-dependent Raman responses as a function of both incident and scattered light polarization.

**2.2 Angle-Resolved Raman Spectroscopy (AARS) Instrumentation**

For this study, the ARRS measurements were obtained using a customized LabRAM HR Evolution confocal Raman microscope (HORIBA France SAS). Excitation involves one of four laser wavelengths ($\lambda$ = 473 nm, 532 nm, 632.82 nm gas laser (abbreviated to 633 nm), or 785 nm) focused through an objective, with the scattered light collected in a 180° back scatter geometry. The objective used in the main body of the paper has 50× magnification with a numerical aperture of 0.75 NA. The confocal hole is fixed to 100 μm for all experiments. Polarization optics include a set of linear polarizers (03 FPG 003 Melles Griot Dichroic Sheet Polarizers) (P1, P2) and two superachromatic half-wave plates (Thorlabs, SAHWP05M-700) (HWP1, HWP2) for linear polarization. In the case of the simplest helicity-resolved polarization measurements discussed here, two linear polarizers (P1, P2), one superachromatic half-wave plate (HWP2), and one superachromatic quarter-wave plate (Thorlabs, SAQWP05M-700) (QWP) are used. The waveplates are held in custom-built motorized mounts capable of a minimum 0.5° step-size. The grating used has 1,800 lines per mm.

As shown in **Figure 1a**, the first optic along the beam path is a broadband, minimal reflective, polarizer (P1). This optic is used to set the incident light, $\lambda$. The polarization vector of λ is denoted as $\boldsymbol{E_i}$ throughout the manuscript and $\phi$ represents the relative angle of the incident

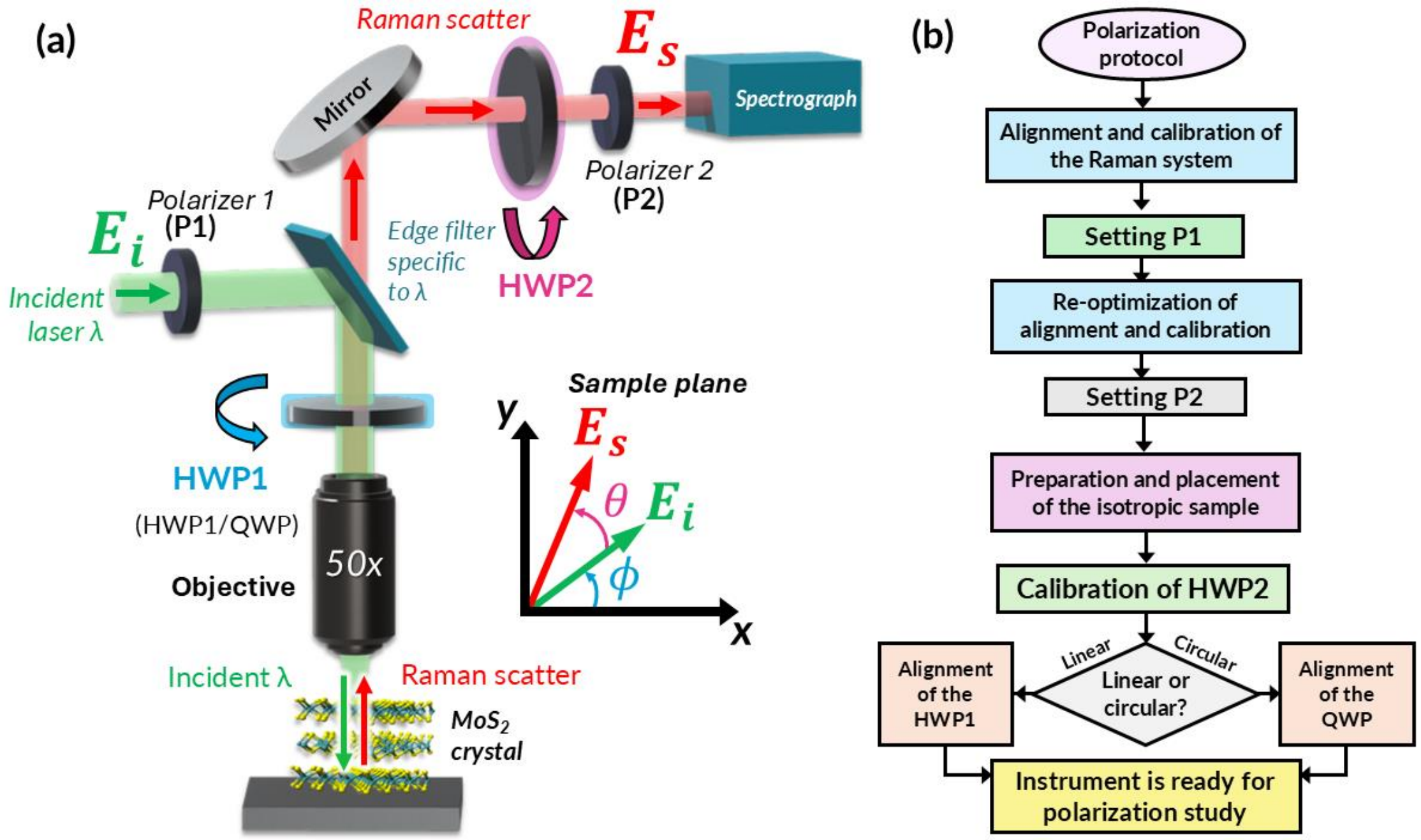

**Figure 1** (a) Schematic diagram of the confocal Raman system for polarization measurements with calibration samples such as molybdenum disulfide ($MoS_2$), where the crystalline *a*-*b* plane is positioned perpendicular to the incident $\lambda$. The angle of rotation between the incident ($E_i$) and scattered ($E_s$) light is denoted as angle $\theta$ in the *x*-*y* laboratory frame (collinear to the a-b plane of the crystal), and the angle $\phi$ is the sample rotation with respect to a fixed $\theta$. Per the inset, the vertical orientation of the incident polarized light is along the x axis. (b) Flowchart describing the required methodology for polarization measurements.

polarization and the orientation in the *x*-*y* plane of the sample. P1 sets along the *x*-direction; this is defined as $\phi = 0°$ (see insert). P1 is also used to minimize the ellipticity of the $\lambda$ at the sample position and was measured to be within a range of ± 4° using a polarimeter (Thorlabs, PAX1000). Next, $\lambda$ interacts with the dichroic edge filter, which is a different optic for each excitation wavelength. The edge filter (angled at 8° to $\lambda$), behaves like a mirror, reflecting the beam toward the sample. Then $\lambda$ is transmitted through a waveplate located at the first motorized mount, between the edge filter and the objective. For linear polarization, this mount holds a half-wave plate (HWP1); for circular polarization, a quarter-wave plate (QWP). (The setup and calibration steps/protocol for circular polarization are discussed in **Section 3.3**.) Note that the location of this first waveplate relative to the edge filter is critical for minimizing edge filter effects such as added ellipticity and polarization distortion.

After the beam passes through the first motorized waveplate, the 50× objective focuses $\lambda$ on to the sample. The resulting Raman backscatter is collected by the objective, passing back through the waveplate (keeping incident and scattered light parallel), and then transmitted through the dichroic filter. Similarly to $\boldsymbol{E_i}$, $\boldsymbol{E_s}$ denotes the polarization vector of the scatter, and $\theta$ represents the relative angle between $\boldsymbol{E_i}$ and $\boldsymbol{E_s}$. At this stage, the edge filter transmits the inelastic Raman scatter and reflects the elastically scattered radiation of the laser line, thus rejecting the Rayleigh scattering from the entrance to the spectrometer. The transmission of Raman scatter through this optic means that the measurable Raman scatter polarization is dependent on the transmission function of the dichroic edge filter. (The role of the dichroic edge filter and its effect on the polarized light will be discussed further in **Section 3.2**.) In the final stage, the Raman scattered beam passes through a second motorized half-wave plate (HWP2) and the other linear polarizer (P2) before entering the spectrograph. The measured Raman intensity depends on two polarization angles $(\theta, \phi)$ in the *x-y* plane of the sample (**Figure 1a** inset); the use of these two different angles is ubiquitous in ARRS measurements.

To experimentally control these angles, various configurations involving different combinations of polarizers and waveplates can be used. The angles $\theta$ and $\phi$ are polarization orientations controlled by the motorized half-waveplates, HWP2 and HWP1, respectively. The $\boldsymbol{E_s}$ polarization is measured using the HWP2 with P2, both of which effectively function as the polarization analyzer found in typical Raman setups. Varying the HWP1 is equivalent to rotating the sample, i.e. fixed polarization with sample on rotation stage. This is because when the reflecting/scattering beam passes through the HWP1 the second time, the waveplate effect is "canceled" out. The accuracies of the azimuth and ellipticity for the HWP optics were fully characterized with several laser wavelengths (with respect to fixed P1 and P2) using a polarimeter.

As mentioned previously, the measured ellipticity of the HWPs is within ± 4° for all tested wavelengths in the configuration shown in **Figure 1a**. By testing two experimental configurations, we demonstrate that the placement of the HWP has a significant impact on polarization accuracy and ellipticity, as detailed in **Figures S2** and **S3**. To further investigate the optical behavior along the detection path, we used an internal diode laser (**Figure S3**), provided by the manufacturer, to precisely align the beam path backwards from the spectrograph through the confocal hole, dichroic filter, and to the sample position. This diode laser, combined with a polarimeter, was used to characterize HWP2. The next section presents a step-by-step protocol for aligning and calibrating the various optical components introduced here to ensure accurate ARRS measurements.

### 2.3 Protocol for accurate Angle-Resolved Raman Spectroscopy (ARRS) measurements

**Figure 1b** presents a flowchart outlining the protocol implemented to establish accurate ARRS measurements. The protocol begins with the major laser alignment of the Raman spectrometer without any polarization optics. Most commercial confocal Raman instruments offer some version of laser alignment and calibration, which includes optimal laser positioning, i.e., the incident angle of the laser to the dichroic edge filter for each excitation laser line. Following these procedures ensures the correct operation and alignment of the input lasers from the source to the sample stage and then from the sample stage to the spectrograph.

The laser alignment in our Raman system involves controlling the beam path using two mirrors for each excitation laser. The laser beam is focused through a microscope objective to a diffraction-limited spot on a Si(111) sample. When well aligned, the spot (approximately 1 μm in size) expands and contracts symmetrically through the focus (along the z-axis) and does not "walk" asymmetrically from right to left or up to down. The laser spot size is also constant (assuming the sample is flat) during *x-y* movement of the sample stage.[29] In addition, we measure the Si $T_{2g}$ phonon intensity as a function of the confocal hole opening. To determine if the system is aligned correctly, we compare the Raman signal collected with the 50 μm and 200 μm sizes. The ratio between the two measurements should be above 80% as recommended by the manufacturer.

Following the laser alignment procedure is the spectral calibration of the instrument, a process repeated at regular intervals. We use a neon lamp to calibrate and verify the frequency wavelength (nm) of the spectrograph against the spectral lines in the NIST Atomic Spectra Lines Database.[30] Additionally, before any Raman experiment, spectral calibration using a doped crystalline Si(111) wafer is performed to verify the Raman shift ($cm^{-1}$) of the spectrum by marking the wavenumber position of the $T_{2g}$ phonon (observed at 520.89 $cm^{-1}$) at its maximum intensity (counts per second). The spectral values obtained during these steps acts as benchmark reference values in terms of the basic instrument function and normal response of the confocal Raman instrument.

The spectral response function of the instrument is also critically important, i.e. the instrument may be more or less sensitive at certain frequencies than at others, as seen recently when using ratios of peaks to quantify defects in graphene.[31] Many manufacturers have a function for performing this calibration that utilizes either a standard broadband light source, or a laser laser-line-specific NIST Standard Reference Material (SRM).[32,33] All instruments need to have a correction to flatten the response over large spectral windows.

After aligning and calibrating our Raman system without polarization optics the polarization protocol (**Figure 1b**) begins with the setting/placement of P1 in **Figure 1a**. Polarizer P1 is inserted into the $\lambda$ input pathway before the dichroic edge filter, and a polarimeter is placed at the sample position. The polarimeter measures the polarization of the light, and its ellipticity. P1 is then rotated to minimize the ellipticity by optimizing the position of the optical fast axis such that $\phi = 0°$. By design, our lasers send in polarized light along the *x*-axis into the Raman instrument. For some arbitrary laser polarization, some combination of HWP and polarizer might be necessary to select the "correct" polarization and minimize the ellipticity at the sample. After the insertion of P1, a small shift in the laser focus position might occur. To rectify this, the laser alignment and spectral calibration process must be repeated to ensure the optimal beam path for the Raman scatter to enter the spectrograph. As an alternative method, P1 can also be aligned without the use of a polarimeter using a wired grid polarizer on a glass substrate as a reference (**Figure S4**). This method relies on minimizing transmitted power through the reference polarizer to determine the correct orientation of P1 along the *x*-axis.

With P1 correctly aligned and the instrument alignment optimized as described above, the next step in the protocol is to place the second polarizer, P2. The point of placing P2 is to select the polarization that corresponds to the highest sensitivity of the spectrograph. To achieve that, the fast axis orientation of P2 can be optimized to the highest polarization response of the spectrometer using the Raman intensity from Si(111). P2 is rotated to maximize the Raman response of the Si $T_{2g}$ phonon, which is isotropic in the (111) plane, for any of the excitation wavelengths available. For our measurements, the highest polarization response of our spectrometer (as set by the manufacturer) corresponds to the vertical orientation of the incident $\lambda$.

The next step is to prepare and place the test material into the sample position, ensuring a flat, clean surface to maximize the Raman signal. The condition of the surface is essential for precise polarization measurements; thus, the test materials must be flat, relatively free of debris and not environmentally degraded. Following our flowchart in **Figure 1b**, we first consider the calibration of the HWP2, and then HWP1 will be discussed. Our protocol introduces the use of test materials such as bulk 2H-$MoS_2$ or c-cut $Al_2O_3$, which are crystal cuts with isotropic Raman response, widely available, and possess well-characterized, in-plane polarization-dependent ($A_{1g}$) and polarization-independent ($E_{2g}$, $E_g$) phonon modes. To calibrate HWP2, we record the $A_{1g}$ response of these crystals as a function of $\theta$ (rotation of HWP2), revealing their characteristic polarization-dependent behavior, as demonstrated with $MoS_2$**.**

Rotating HWP2 in 5-degree increments and plotting the area of the $MoS_2$ Raman peaks (**Figure S5**) shows that the maximum area of the $I_{A_{1g}}$ corresponding to the Raman scattering aligned with the fast-axis of the P2 optic, where $\boldsymbol{E_i}$ and $\boldsymbol{E_s}$ are parallel along the *x*-axis. We define the polarization of $\boldsymbol{E_i}$ and $\boldsymbol{E_s}$ as $\boldsymbol{k_s}(\boldsymbol{E_s}\boldsymbol{E_i})\boldsymbol{k_i}$ (the Porto notation), where $\boldsymbol{k_i}$ and $\boldsymbol{k_s}$ are the incident and scattered propagation directions. Using the Porto notation, this condition determines parallel ($\boldsymbol{z}(\boldsymbol{xx})\boldsymbol{\bar{z}}$) polarization, which we define as $\theta = 0^{\circ}$. The corresponding minima of the $I_{A_{1g}}$ is the perpendicular/crossed ($\boldsymbol{z}(\boldsymbol{yx})\boldsymbol{\bar{z}}$) polarization and defined as $\theta = 90^{\circ}$. In the same plot we can also see that the $E_{2g}$ mode at 384.25 $cm^{-1}$ in $MoS_2$ is not polarization-dependent, *i.e.*, is a flat line, exhibiting no obvious change in their respective intensities as a function of HWP2 rotation.

Having aligned the HWP2, the last step in the protocol for linear polarization is the calibration of HWP1. The HWP1 is placed above the objective, and the test material is replaced by the polarimeter. The HWP1 is rotated such that its initial position ($\phi$ = 0°) corresponds to $\boldsymbol{E_i}$ along the *x*-axis ($\boldsymbol{E_i} \parallel \boldsymbol{x}$). From our tests, this alignment can be carried out using a polarimeter, or an alternative method that uses another polarizer and a power meter (**Figure S6**). Note that for circular polarization, a QWP is used in place of HWP1. The QWP alignment is detailed in the helicity-resolved Raman polarization section (**Section 3.3**).

With the selected polarization optics optimized for linear or circular polarization, the motorized control of the HWPs enables the capability to vary both $\theta$ and $\phi$ independently to generate 2D polarization maps (2D-PM), which is described in the following section, along with the results and discussion for materials with isotropic and anisotropic phonon response.

## 3. Results and Discussion

### 3.1 Linear polarization: Materials with an Isotropic Raman Response

We first investigate the angle-resolved Raman response from a bulk 2H-$MoS_2$ sample, which was prepared as described in Section 2.1. **Figure 2a** shows the room-temperature Raman spectra from $MoS_2$ measured with four distinct excitation wavelengths (473 nm, 532 nm, 633 nm, and 785 nm). The spectra exhibit two strong Raman peaks corresponding to the and the in-plane $E_{2g}^1$ (384.25 $cm^{-1}$) and out-of-plane $A_{1g}$ (409.27 $cm^{-1}$) optical vibrational modes.[34–36] The insets at the top of **Figure 2a** provide schematics describing the relative displacements of S and Mo atoms in a 2H-phase $MoS_2$ crystal for the $E_{2g}^1$ and $A_{1g}$ modes. **Figure 2a** also shows that the experimental data obtained with 633 nm excitation (red curve) differs from other excitation wavelengths. This is due to 633 nm light being nearly resonant with both the A exciton (655 nm) and the B exciton (610 nm) in $MoS_2$.[26,37] With the resonance process, when the excitation laser is located near an absorption of the materials, several second-order Raman modes are distinctly observed above the noise level as seen in the red spectrum. We suggest that users avoid 633 nm excitation for polarization calibrations using $MoS_2$.

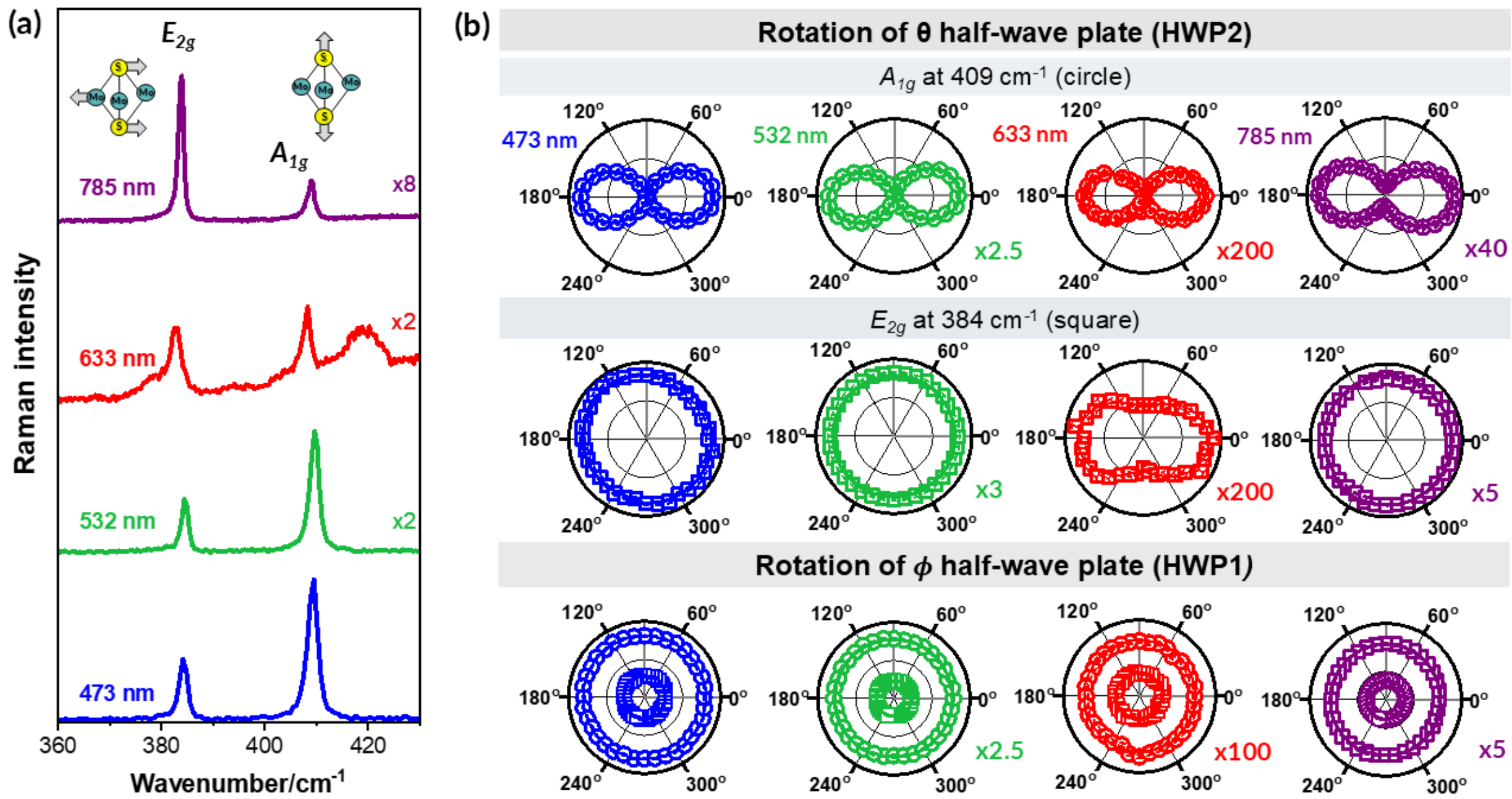

**Figure 2.** (a) Room-temperature Raman spectra from bulk $MoS_2$ at four distinct excitation wavelengths showing the in-plane $E_{2g}$ (384.25 cm$^{-1}$) and out-of-plane $A_{1g}$ (409.27 cm$^{-1}$) vibrational modes collected with a 50× objective. The 633 nm is the resonant Raman response of $MoS_2$. (b) Polar plots of the integrated area (Lorentzian fitting) of the $I_{A_{1g}}$ (open circles) and $I_{E_{2g}}$ (open squares) peaks are plotted as a function of the two different half-wave plate rotation, revealing their complementary polarization characteristics. Note the sizeable differences in scale.

To examine the polarization response of these modes and validate our protocol, we analyze the Raman response of these phonons while varying both $\theta$ and $\phi$ as seen in **Figure 2b.** For $\theta$ rotation, the incident polarization is fixed along the *x*-axis, $\phi = 0°$, however in bulk 2H-$MoS_2$ or c-cut $Al_2O_3$ samples, the incident polarization can be oriented randomly while still obtaining the same result. On the other hand, for $\phi$ rotation, $\theta$ is fixed at zero for the maximum response of the $A_{1g}$ mode, corresponding to the parallel configuration. For the $E_{2g}^1$ mode, any position of $\theta$ will give the same result. **Figure 2b** summarizes the polar plots obtained from the integrated Lorentzian peak areas of the $A_{1g}$ and $E_{2g}^1$ modes as functions of the angular rotations, $\theta$ and $\phi$, for four excitation lasers.

By varying $\theta$— (**Figure 2b** top and middle panels) that is, sweeping the angle between $\boldsymbol{E_i}$ and $\boldsymbol{E_s}$ from 0° to 360°— the $A_{1g}$ and $E_{2g}^1$ modes of $MoS_2$ have distinct polarization responses. The magnitude and differences in the polarized Raman scattering intensity between the $A_{1g}$ and $E_{2g}^1$ vibrational modes derive from the differences in their respective Raman tensors. The intensity of the $A_{1g}$ mode has a response proportional to $\cos^2(\theta)$ producing a polar plot with a two-fold symmetric 'peanut' shape. The polar plots in the top row of **Figure 2b** show the experimental data along with their fitting to the function $\cos^2(\theta)$. From our observations, the intensity of the two lobes of the 'peanut' can be affected by misalignment, ellipticity, and/or retardation of light by optics in the beam path. If the lobes are asymmetric, one should be concerned about the quality of the polarization measurements. The maximum integrated area corresponds to $\theta = 0°$ (angle between the $E_i$ and $E_s$) where the polarization of the incident $\lambda$ and scattered $\lambda$ are parallel. This is not surprising as our polarization protocol described in **Section 2.3**, designates the HWP2 position at maximum $I_{A_{1g}}$ as the $\theta = 0^{\circ}$, defined as parallel ($VV$).

Conversely, the minimum of $I_{A_{1g}}$ occurs when the polarization of the incident $\lambda$ and scattered $\lambda$ are perpendicular to each other at $\theta = 90^{\circ}$or perpendicular ($VH$). In contrast to the $A_{1g}$ mode, the doubly degenerate, in-plane $E_{2g}^1$ vibration does not have a $\theta$ dependent polarization response [$(I_{E_{2g}^1} \propto \cos^2(\theta) + \sin^2(\theta) = 1)$], and therefore no overall peak maximum is observed, thus the $I_{E_{2g}^1}$ remains constant. We observed in the middle row of **Figure 2b** that the unchanging intensity (circular plot) of the $E_{2g}^1$ mode is indicative of the stability of the instrument and confirms the correct alignment of the excitation laser. However, the polar plot of the $E_{2g}^1$ mode with 633 nm excitation differs from other excitation wavelengths and deviates from the ideal response because the intensity of the second-order modes inhibits quality fits of the first-order $A_{1g}$ and $E_{2g}^1$ Raman peaks and thereby alters their plotted polarization dependence. This deviation from the ideal polarization response under 633 nm excitation has been reported in $MoS_2$ before,[38] and points to the need to consider resonance conditions when performing accurate ARRS calibration or measurements. For polarization measurements when only $\lambda$ = 633 nm is available, we recommend using the $A_{1g}$ mode of a different test material, $Al_2O_3$(001) as the calibrant, as its Raman peaks are not resonant at that specific wavelength. While $Al_2O_3$(001) is more robust under environmental exposure and high laser power, its lower signal-to-noise ratio requires longer acquisition times and higher excitation power, making it a less efficient alternative to $MoS_2$ for calibration purposes (**Figure S7**).

Having discussed $\theta$ rotation, we now investigate the response of $MoS_2$ Raman active modes in response to $\phi$ rotation. For bulk 2H-$MoS_2$ or c-cut $Al_2O_3$, the relative angle of polarization of the incident light to the crystal axis ($\phi$) will not affect the phonon modes. The bottom row of **Figure 2b** shows the Raman response of the $A_{1g}$ and $E_{2g}^1$ modes in $MoS_2$ as the incident polarization is rotated along the crystallographic *ab*-plane $0^{\circ} \leq \phi \leq 360^{\circ}$. Given that this is analogous to an in-plane sample rotation, the absence of variation in the polarization response verifies the in-plane isotropy of $MoS_2$ phonon modes. The resulting constant intensities for both Raman modes under rotation indicate an absence of any induced ellipticity effect as the HWP1 is rotated. The same data collection procedure was repeated using a 100× objective (**Figure S8**) and is consistent with the 50× objective data.

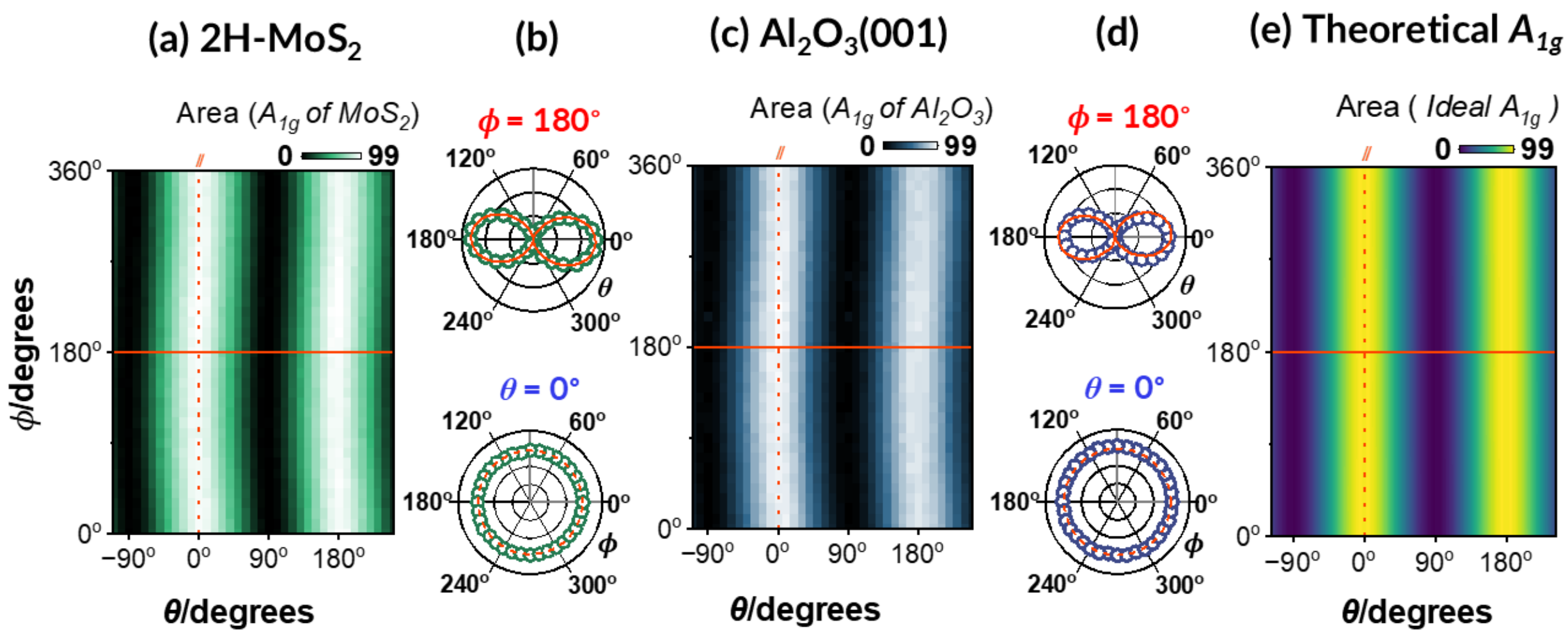

**Figure 3.** 2D polarization maps of the $A_{1g}$ peak area as a function of $\phi$ and $\theta$ in (a) Bulk crystal $MoS_2$ and (c) $Al_2O_3$(001) is consistent with the (e) theoretical prediction of an $A_{1g}$ Raman tensor (Equation 3). Next to the experimental 2D-PMs are the polar plot data (b and d) at $\phi$ = 180° (solid, top) and $\theta$ = 0° (dotted, bottom) (a and c) with an overlay of the theoretical prediction from (e).

In order to gain a more comprehensive understanding of the instrument and sample polarization, it is useful to collect the polarization response of the Raman modes as both angles $\phi$ and $\theta$ are rotated. Using bulk 2H-$MoS_2$ (**Figure 3a**) and $Al_2O_3$ (001) (**Figure 3c**), we measure the linear polarization response of $I_{A_{1g}}$ with 532 nm excitation laser, as HWP1 and HWP2 are rotated to generate a 2D-PM. The 2D-PMs have their angular rotations, $\theta$ and $\phi$, plotted along the $x$ and $y$ axes, respectively. The plots for both $MoS_2$ and $Al_2O_3$(001) feature two vertical iso-intensity bands with maximum values separated by 180°. As expected from our calibration protocol, the maxima of these bands are located at 0° and 180°, which correspond to the parallel polarization measurement configuration. Extracting the data from the horizontal (solid lines) and vertical (dash lines) slices of the 2D map produces the $A_{1g}$ polar plots presented in **Figure 3b** and **3d.** The top polar plots show the two-fold symmetric "peanut-shape" under $\theta$ rotation, and the bottom polar plots show the invariant (circular) response under sample rotation $\phi$ for each crystal.

For greater insight into the polarization response of our optics, we compare our experimental data with the theoretical polarization response of the $A_{1g}$ phonon. Since the $MoS_2$ Raman tensor modes are known, we can calculate the theoretical polarized Raman response and compare with our experimental results (**Figure 3**). In general, the intensity of the polarized Raman scatter ($\boldsymbol{I_S}$) is described by the equation below.

$$\boldsymbol{I_S} \propto |\boldsymbol{E_s} \cdot \boldsymbol{R} \cdot \boldsymbol{E_i}|^2 \qquad (1)$$

where $\boldsymbol{E_i}$ and $\boldsymbol{E_s}$ corresponds to the polarization vectors of the incident and scattered light and $\boldsymbol{R}$ is the Raman tensor describing the response of a particular vibrational mode. In the case of $MoS_2$, the $A_{1g}$ and $E_{2g}^1$ Raman tensors are:

$$\boldsymbol{R}(A_{1g}) = \begin{bmatrix} a & 0 & 0 \\ 0 & a & 0 \\ 0 & 0 & b \end{bmatrix} \text{ and } \boldsymbol{R}(E_{2g}^1) = \begin{bmatrix} c & 0 & 0 \\ 0 & -c & 0 \\ 0 & 0 & 0 \end{bmatrix}, \begin{bmatrix} 0 & c & 0 \\ c & 0 & 0 \\ 0 & 0 & 0 \end{bmatrix}, \qquad (2)$$

where $a$, $b$, and $c$ are the tensor elements. The basis for these tensors is the Cartesian coordinate of the sample frame, with the *z*-axis being parallel to the out of plane, crystallographic *c*-axis. Adding to **Equation (1)** the typical linear transfer matrices (Jones Matrix formalism[39,40]) for polarizing optics along with the Raman response tensor prescribed by group theory[41,42] in **Equation (2)**, we model the Raman intensity of $A_{1g}$ mode as:

$$I_{A_{1g}}(\theta,\phi) \propto \left|\boldsymbol{E}_s \cdot \boldsymbol{M}_{HWP}(-\theta/2) \cdot \boldsymbol{M}_{HWP}(+\phi/2) \cdot \boldsymbol{R}(A_{1g}) \cdot \boldsymbol{M}_{HWP}(+\phi/2) \cdot \boldsymbol{E}_i\right|^2 \qquad (3)$$

where, $\boldsymbol{M}_{HWP}(\alpha)$ corresponds to the Jones Matrix representation of a half-wave plate that is physically rotated through an angle $\alpha$ in the instrument [see **Supplemental Equation (S1)**]. Note that the signs and magnitudes of the angles specified in **Equation (3)** correspond to the directions of incident ($\phi$) and scattered ($\theta$) light polarization as shown in **Figure 1a**; specifically in our instrument, HWP1 rotates counterclockwise ($+\phi/2$) and HWP2 rotates clockwise ($-\theta/2$). The last $\boldsymbol{M}_{HWP}$ term contains a minus sign indicating that the second half-wave plate rotates in the opposite sense of the first half-wave plate, in order to correct for the rotation imposed on the scattered beam propagating back through HWP1. Using the Raman tensors in **Eq. (2)** and **Eq. (3)** for each of the $MoS_2$ modes, the polarization responses are calculated and compared with the experimental results for $\theta$ or $\phi$ rotations. As can be seen in **Figure 3e**, the simulated response of the $A_{1g}$ phonon in a material that produces an isotropic polarization response matches well with the experimental 2D polarization maps from $A_{1g}$ phonons in both $MoS_2$ and $Al_2O_3$(001). Unlike the more complex and anisotropic $\boldsymbol{R}(B_{3g})$ tensor modeled by Wood *et al.*[3], the simpler $\boldsymbol{R}(A_{1g})$ **Equation (2)** produces an $\boldsymbol{E}_s$ polarization whose instrumental factors are symmetric with respect to the *x-y* laboratory frame.

To investigate the effect of numerical aperture (NA) on our polarization-resolved Raman study, we performed ARRS on three materials (2H-$MoS_2$, (100) and (111) cuts of Si) using four different objectives with magnification (and NA) as follows: 10× (0.25), 50× long-working distance (0.5), 50× (0.75), and 100× (0.9). The resulting 2D polarization maps (**Figure S9**) show negligible influence from non-normal incident light ($\boldsymbol{E_i}$ or $\boldsymbol{E_s}$ are polarized along $\boldsymbol{z}$), as discussed in Supplemental Information, **S8**. This result, consistent with Turrell et al.[17], is supported by our modeling that attributes less than 2% of the total response to $\boldsymbol{z}$-polarized light contributions.

### 3.2 Linear polarization: Materials with an Anisotropic Response

Following a thorough analysis of the ARRS measurements on crystalline materials with a model isotropic polarization response, we now apply our methodology and instrumentation to materials with an anisotropic  polarization response. We select silicon as the test material for this section, specifically the anisotropic response of a $T_{2g}$ phonon in Si(100). Single crystal Si forms a diamond structure (cubic space group $\mathrm{Fd\bar{3}m}$) with a triply-degenerate $T_{2g}$ Raman-active phonon at 520.89 $\mathrm{cm^{-1}}$, whose Raman tensor gives rise to a $\phi$-dependent intensity in the (100) scattering plane.[3] **Figure 4a** shows the experimental polarization response of the phonon $T_{2g}$ with excitation $\lambda$ = 532 nm. Note that the experimental 2D-PM has a "checkerboard" appearance indicating that the maxima (white) and minima (black) are periodically separated, showing the $T_{2g}$ symmetries: 4-fold symmetric (90°) with $\phi$ rotation, and 2-fold symmetric (180°) with $\theta$ rotation. To facilitate the comparison between experimental and theoretical responses, we selected two pairs of horizontal and vertical slices (red and cyan lines) from the 2D-PMs to produce polar plots for $\theta$ and $\phi$ rotations, at specific angles shown in **Figures 4b** and **4c**, respectively.

Per our methodology, the polar plots are obtained from the integrated Lorentzian peak area of the $T_{2g}$ mode as functions of the angular rotations. The polar plots resulting from the first pair of slices (red) are shown in the top of **Figure 4b** and **4c**. The top plot in **Figure 4b** shows the $T_{2g}$ response as a function of $\theta$, at $\phi$ =180°, indicating that the $\boldsymbol{E_i}$ is fixed ($\boldsymbol{E_i} \parallel \boldsymbol{x}$). The corresponding slice in the 2D-PM is marked as a red solid horizontal line in **Figure 4a.** The experimental response of the $T_{2g}$ mode (diamonds) under $\theta$ rotation is as expected to be minimized for $\theta$ = 90° (cross polarized) and maximized for $\theta$ = 0° (parallel). Looking at the 2D-PM in **Figure 4a**, slices at fixed $\phi$ produce two-fold symmetric, "peanut" shape polar plots, e.g., as in **Figure 4b**.  If we extract a slice of the map fixed at $\phi$ = 45° the "peanut" shape is rotated, resulting in a vertical "peanut" shape with maximum at $\theta = 90^{\circ}$. This "peanut" shape rotates when polarization of the incident light is rotated in the (100) plane.

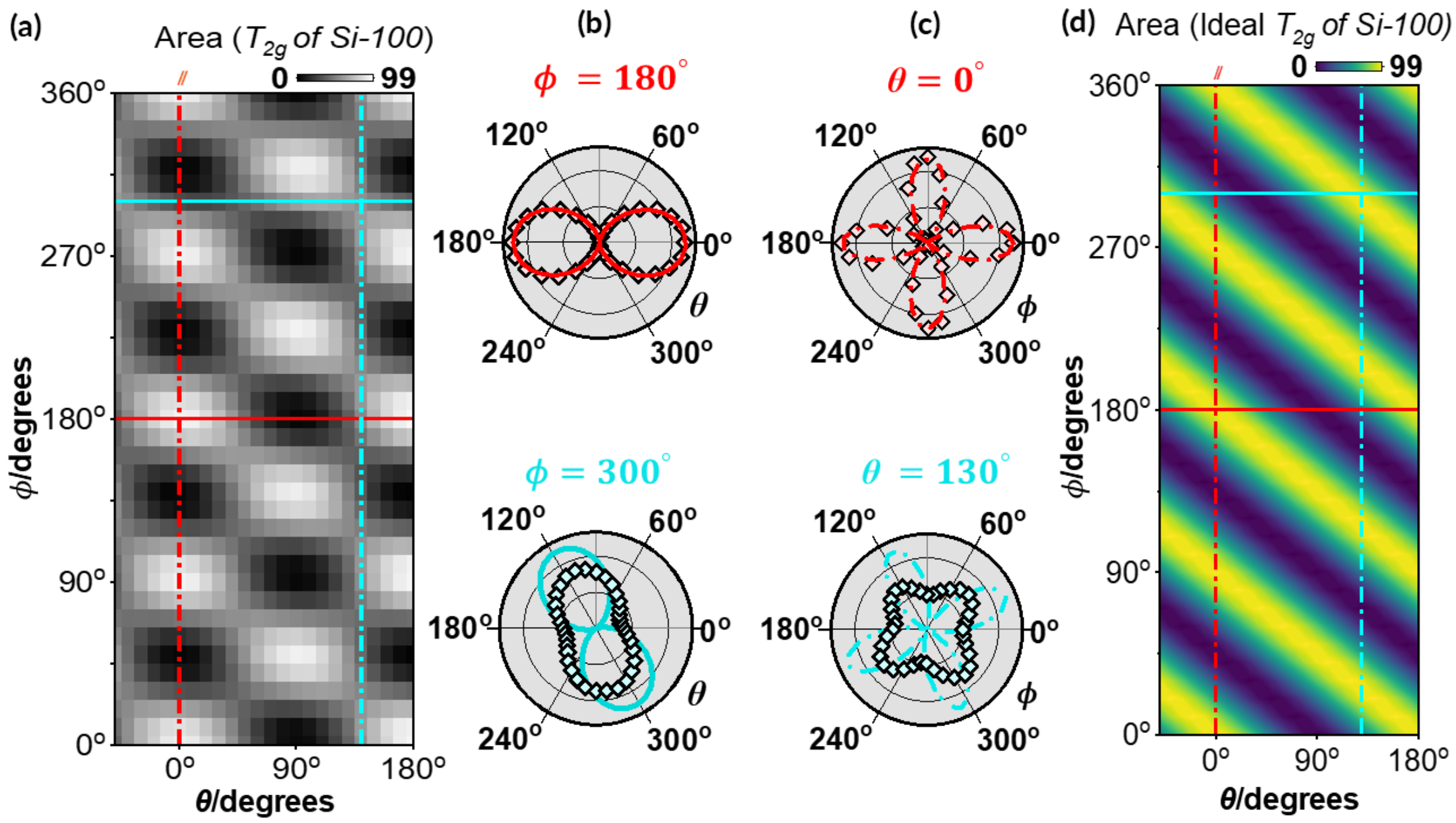

**Figure 4.** 2D-PM of the $I_{T_{2g}}$ peak area as a function of the $\phi$ and $\theta$ in anisotropic Si(100): (a) The measured polarization map shows differences from the (d) calculated Raman $T_{2g}$ tensor. (b) Polar plots at fixed $\phi$ = 180° and $\phi$ = 300° (diamonds) with an overlay of the theoretical prediction (solid) as a function of θ rotation. (c) Polar plots at fixed θ = 0° and θ = 130° (diamonds) with an overlay of the theoretical prediction (dashed) as a function of $\phi$ rotation. Unlike the more uniform isotropic polarization maps, the experimental $T_{2g}$ does not overlap with the theory at periodic angles, such as $\phi$ = 300° and θ = 130°.

The experimental results of the Raman response of the $T_{2g}$ phonon in Si(100) confirm the correlation between the incident polarization, and the crystal plane orientation. The top polar plot in **Figure 4c** displays the response for $\phi$ rotation under the parallel ($VV$) scattering configuration ($\theta = 0^{\circ}$) and corresponds to the red dashed vertical line in **Figure 4a**. The "flower-shape" of this plot confirms the anisotropy and four-fold symmetry of the $T_{2g}$ mode in Si(100).[43,44] The bottom polar plots of **Figures 4b** and **4c** show the Raman response of $T_{2g}$ mode for the other pair of slices (cyan lines) in the 2D-PM. The bottom polar plot of **Figure 4b** corresponds to the $T_{2g}$ polarization response under $\theta$ rotation at fixed incident polarization $\phi$ = 300°. The bottom polar plot of **Figure 4c** displays the response for $\phi$ rotation, with the incident and Raman scattered polarization vectors not fully parallel ($\theta$ = 130°). Compared to the first (top) pair, the experimental results for the second (bottom) pair differ from the expected "peanut" and "flower" shapes.

To better understand the origin of these unexpected shapes, we calculate the theoretical response of the $T_{2g}$ mode in Si(100). The $I_{T_{2g}}$ phonon response is the sum of the individual intensities of the tensors for the $T_{2g\,(x)}$, $T_{2g\,(y)}$, and $T_{2g\,(z)}$ degenerate modes, i.e.,

$$\boldsymbol{R}\big(T_{2g(x)}\big) = \begin{bmatrix} 0 & 0 & 0 \\ 0 & 0 & d \\ 0 & d & 0 \end{bmatrix};\ \boldsymbol{R}\big(T_{2g(y)}\big) = \begin{bmatrix} 0 & 0 & d \\ 0 & 0 & 0 \\ d & 0 & 0 \end{bmatrix};\ \boldsymbol{R}\big(A_{2g(z)}\big) = \begin{bmatrix} 0 & d & 0 \\ d & 0 & 0 \\ 0 & 0 & 0 \end{bmatrix};$$

$$I_{T_{2g}} = I_{T_{2g}(x)} + I_{T_{2g}(y)} + I_{T_{2g}(z)}. \qquad (4)$$

In the case of the wafer Si(100), only $I_{T_{2g}(z)}$ survives according to our configuration $\boldsymbol{z}(\boldsymbol{E_s E_i})\bar{\boldsymbol{z}}$, where the incident and scattered light propagate only along the $z$-axis. Thus, in the same way the polarized Raman response was calculated for the $A_{1g}$ mode in **Equation (3)**, we calculate the ideal instrument response for $T_{2g}$ phonons polarization from a Si(100) wafer as:

$$I_{T_{2g}}(\theta, \phi) \propto |\sin[\theta + 2(\phi - \phi_o)]|^2, \qquad (5)$$

where $\phi_o$ indicates the relative incident polarization $\lambda$ respect to the in-plane crystallographic axis. The $\phi_o = -\frac{\pi}{4}$ is based on the sample position and methods (see **Section 2.1**). Note **Equation (5)** is for the angles defined in **Figure 1a** and corresponds to HWP2 rotating clockwise (CW) observed looking along the z-axis from the detector. If both HWPs rotate in the same relative direction, the argument for the sine function changes to $[\theta - 2(\phi - \phi_o)]$ in **Equation (5)**.

Using **Eq. (5)**, we calculate the theoretical 2D-PM response as a function of both the $\theta$ and $\phi$ angles (**Figure 4d**). The ideal 2D-PM has a diagonal stripe pattern, which differs from the experimental one. As we can see the $\theta$-$\phi$ contour plot shows diagonal iso-intensity bands. The bands slope with a sign dependent on the rotation direction of the HWPs. For our instrument configuration, the diagonal bands have a negative slope corresponding to **Eq. (5)**. As seen in **Figure 4d**, the predicted pattern indicates no expected distortion of the 2-fold symmetric "peanut shape" nor of the 4-fold symmetric "flower shape", but rather only a rotation of the symmetric axis as a function of $\theta$ and $\phi$, respectively. For comparison, **Figure S10 s**ummarizes the 2D experimental polarization maps and ideal response simulation of the $T_{2g}$ peak area for two other cuts of silicon. As expected for Si(111), there is congruence between simulation and experimental data, however, the anisotropic response from Si(110) also shows discrepancy with the ideal response simulation. Below we focus only on analysis of the Si(100) cut.

From the theoretical 2D-PM of Si(100), the same pair of slices were extracted and overlayed with the experimental points in the polar plots in **Figure 4b** and **4c.** We confirm the agreement of the first pair of slices (red), and the non-congruency for the second pair (cyan) with the theoretical response. We notice that the response at the specific values of $\phi$ = 300º and $\theta$ = 130º, demonstrates a discrepancy between the experimental and ideal responses, showing distorted polarization response of the $T_{2g}$ phonon mode. Furthermore, comparing the theoretical response in **Figure 4d** with the 2D-PM in **Figure 4a** we see that the periodic, non-congruent $T_{2g}$ response ("checkerboard" pattern) at angles beyond the specific parallel $(VV)$ and perpendicular $(VH)$ polarization configurations indicate contribution(s) of external factor(s) not modeled in **Eq.**

**(5)**. While the experimental results of materials with an isotropic response ($MoS_2$, $Al_2O_3$) are corroborated by the predicted 2D-PMs, materials with an anisotropic response are significantly more sensitive to issues with the instrument response. By using a polarimeter to characterize the polarization response of the optics in the ARRS setup (**Figure S2**), under different configurations and along different light paths, we identified the contribution of the dichroic edge filter as causing a significant effect on the polarization of the Raman scattering from materials with an anisotropic response.

To effectively target the instrumental depolarization effects observed in ARRS systems, we now consider the effects from the convolution of the dichroic edge filter and the intervening optics effects (two flat mirrors) in the theoretical calculation of the $I_{T_{2g}}(\theta, \phi)$. Based on our optics characterization and the polarization compensating arrangement of the two mirrors, the major contribution in the depolarization effects is from the dichroic edge filters (**Figure S2**). In the selected ARRS configuration, the dichroic edge filters, with high reflectance at the laser line and high transmittance of the Stokes-shifted Raman signal, are used to filter the excitation source wavelength from the emission wavelengths. While the excitation light is polarized in the vertical direction, corresponding to $s-$polarization on the filter, the Raman emission of materials can be a mixture of $s-$ and $p-$polarized light. Therefore, we model the dichroic optic using a transfer matrix with arbitrary attenuations and phase shifts for the $s$- and $p$-polarizations.

The schematic in **Figure 5a** provides a brief overview of the nature of dichroic filters in altering the polarization output. We introduce another matrix $\boldsymbol{M}_{edge}$ to represent the non-ideal property of the edge filter as a generic waveplate with some arbitrary phase shift and transmission coefficients:

$$\boldsymbol{M}_{edge} = \begin{bmatrix} t_s & 0 \\ 0 & t_p * e^{i\delta} \end{bmatrix} \qquad (6)$$

where $t_s$ and $t_p$ are the amplitudes of the complex Fresnel transmission coefficients, respectively, and $\delta$ is the relative phase difference. Inserting the filter matrix into the scattering path in **Eq. (5)** gives

$$I_{T_{2g}}(\theta,\phi) \propto \left|\boldsymbol{E}_s \cdot \boldsymbol{M}_{HWP}(-\theta) \cdot \boldsymbol{M}_{edge} \cdot \boldsymbol{M}_{HWP}(\phi) \cdot \boldsymbol{R}\left(T_{2g}\right) \cdot \boldsymbol{M}_{HWP}(\phi) \cdot \boldsymbol{E}_i\right|^2. \quad (7)$$

Note that **Equation (7)** only considers the transmission through the optics and negates reflection. We ignore the reflection because the polarization of the incident light is maintained as it reflects off the edge filter of purely $s$ −polarized light due to the calibrated alignment. Also, the physical arrangement of the two mirrors (compensating the $s$- and $p$- polarization reflections) allows us to ignore their contribution.

The resulting expression for the $T_{2g}$ phonon polarization response from a Si(100) wafer using **Eqs. (6)** and **(7)**, accounting for the edge filter effect,[‡‡] is

$$I_{T_{2g}}(\theta, \phi) \propto \left| e^{i\delta} t_p \sin(\theta) \cos(2\phi) + t_s \cos(\theta) \sin(2\phi) \right|^2 . \qquad (8)$$

[‡‡] Note that, generally, there would be an additional $\theta_0$ and $\phi_0$ used when fitting these expressions. $\theta_0$ is set to zero, as part of HWP2 (in the Raman beam path) calibration. $\phi_0$ is determined by the orientation of the Si(100) wafer when placed onto the sample stage. These constant angles can be thought of as systematic offsets due to experimental conditions and have been left out for clarity.

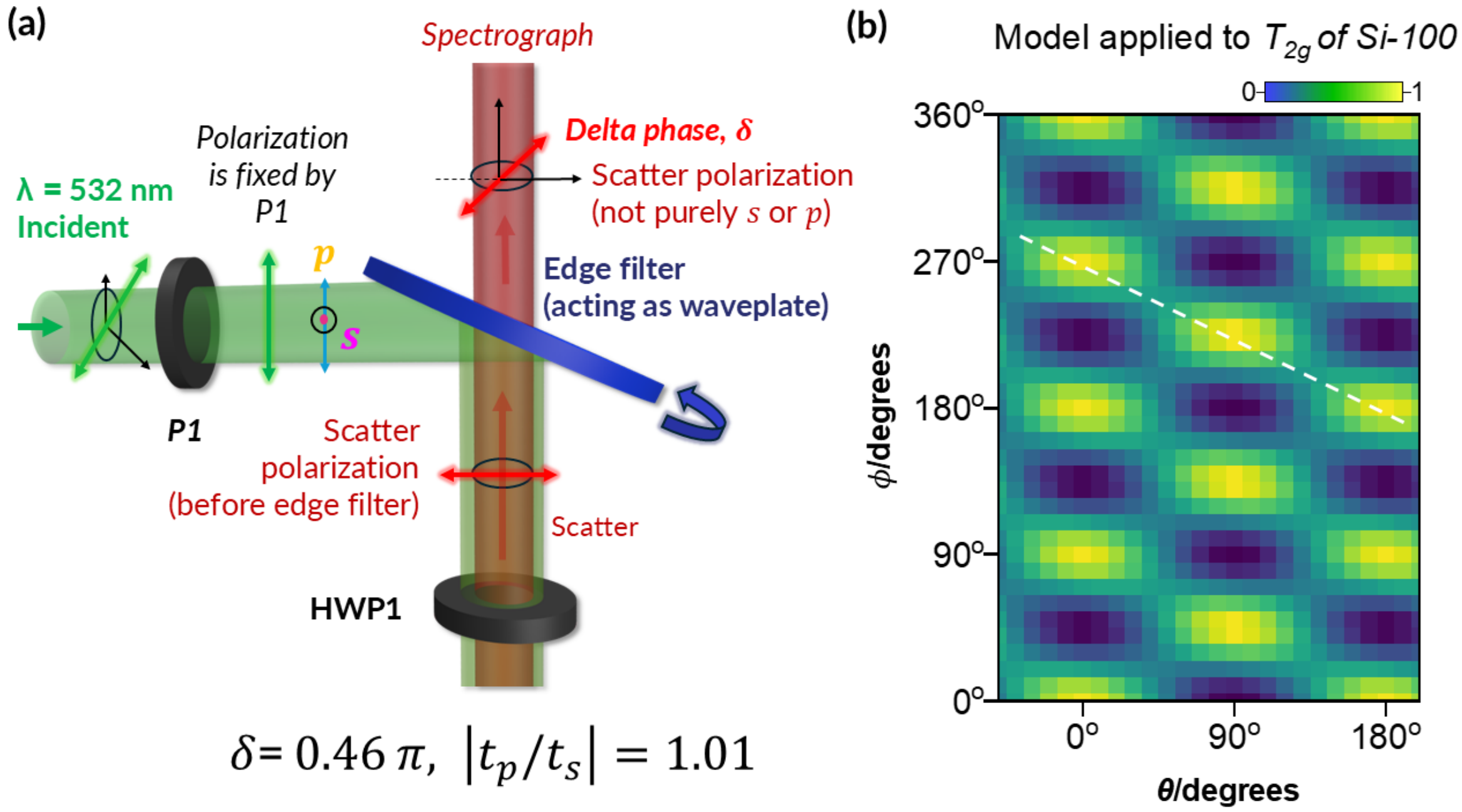

**Figure 5.** (a) Schematic of the polarization optics illustrating the effect from the edge filter inducing a phase response from Raman scatter that is neither purely *s*- nor *p*-polarized. (b) Experimental data of the 2D polarization map for Si(100) (**Figure 4a**) fitted using the model described in Equation (8). Note that the white dashed line guides the eye to identify the slope of the modulated diagonal bands in the pattern.

This theoretical result for the $I_{T_{2g}}$ response generates the diagonal iso-intensity bands in a 2D-PM, replicating the checkerboard pattern in the experimental data for certain values of $\delta$ near $\frac{\pi}{2}$. Our experimental 2D-PM shows that the bands slope towards the negative ($\delta < \frac{\pi}{2}$) as shown by the white dashed line in **Figure 5b**. The impact of $\delta$ and how its value affects the symmetry, or the slope of the bands are shown in **Figure S11**.

Using **Equation (8)**, we fit the experimental data in **Figure 4a** to produce the 2D-PM in **Figure 5b**, which includes the edge filter effect. From fitting the Si(100) data measured with the 532 nm edge filter, we extract an additional phase of $\delta = 0.46\pi$ introduced by the filter. For comparison, a quarter-wave-plate produces a relative phase shift of $\delta = 0.50\pi$ between the fast and slow axes. Each edge filter, tuned to a specific wavelength, induces a distinct phase shift, thereby introducing a polarization distortion, which changes the modulated bands seen in the 2D-PM (**Figure S11)**. For each excitation wavelength, we quantify how well the model reproduces the experimental results using the estimated variance (**Equation S1**). The extracted phase shifts (and estimated variance) from corresponding fittings are: 0.46π (0.0045) for 532 nm, 0.38π (0.019) for 473 nm, 0.51π (0.067) for 633 nm, and 0.62π (0.0024) for 785 nm, as summarized in **Figure S12** and **Table S1**.

Building on this analysis of edge filters matched to specific wavelengths, we extended our study to Volume Bragg Gratings (VBG) filters (a type of notch filter often used for ultra-low frequency measurements) to explore whether similar wavelength-dependent phase shifts might also be present. VBGs are particularly well-suited for this investigation due to their narrow bandwidth filtering (often 5 $cm^{-1}$ to 10 $cm^{-1}$), high rejection efficiency, and excellent transmission characteristics, making them ideal for applications requiring precise wavelength selection and high signal quality.[45] In our study, we used the 532 nm excitation and three VBG filters, including one replacing the edge filter and two positioned along the scattering path, as shown in **Figure S13a**. This setup yielded a 2D-PM consistent with the ideal case (**Figure S13b**). These VBG filters produced significantly less polarization "distortion" than the dielectric long-pass filters. We observed a closer agreement with the simpler model predictions of **Eq. (5)**, and a smaller phase factor, $\delta = 0.10\pi$ (0.0007), when including a filter in the model using **Eq. (8)**.

### 3.3 Helicity-Resolved Raman Spectroscopy (HRRS)

Having discussed the ARRS response with linearly polarized excitation, it is also important to consider helicity-resolved Raman spectroscopy as a technique for studies involving chiral materials.[46] Chiral materials, described as containing symmetries whose mirror images cannot be superimposed onto each other, interact differently with right-handed and left-handed circularly polarized light. Furthermore, the detection and manipulation of chiral phonons in chiral material are increasingly being explored for their potential applications in chiral induced spin selectivity (CISS)[47], spintronics[48] and valleytronics[49]. Only the simplest of the helicity-resolved measurements using the same aforementioned test materials are discussed here and only for completeness.

Using the same configuration described in **Figure 1a**, we can easily transition to helicity-resolved Raman measurements by substituting the half-waveplate (HWP1) located above the objective with a superachromatic quarter-wave waveplate (QWP). We applied the same polarimetric characterization procedures to the QWP as we did for the HWP. Circularly polarized light was observed when the QWP was rotated ±45° relative to the input linear polarization of λ (**Figure S14**). This orientation places the input polarization at 45° to the QWP fast axis, introducing a π/2 phase shift between orthogonal components and resulting circular polarization. In contrast, when the fast or slow axis of the QWP is aligned parallel to the input polarization, the linear vertical polarization (set by polarizer P1) is preserved.

**Figure 6a** shows a schematic of the optical path for helicity resolved measurements. The incoming polarized $\lambda$ passing through the P1 and QWP (set at $\pm 45°$) produces the circularly polarized (CP) incident light. Note that, we are using the convention used by Barron and Buckingham[50], where right-circular ($R\ or\ \sigma_R$) involves a clockwise rotation of the electric field vector when viewed toward the source of the light. Thus, it follows that the left-circular ($L\ or\ \sigma_L$) is counterclockwise rotation. The Raman scattered light from the sample passes through the QWP, edge filter, HWP2, and P2 before entering the spectrograph. In general, the combination of the QWP and P2 can act as a CP analyzer. Here, HWP2, positioned between the QWP and P2, selects either the $R$ or $L$ helicity of the scattering light. Similar to the linear polarization, the selection of the helicity of the incident and scattered light allows for different measurement configurations. When the selected helicity of the scattering light is identical to the incident CP, the configuration is described as co-rotating: $RR$ ($\sigma_R\ \sigma_R$) or $LL$ ($\sigma_L\ \sigma_L$); if the selected helicity of the scattering light and incident CP are opposite, the configuration described as counter-rotating: $RL$ ($\sigma_R\ \sigma_L$) or $LR$ ($\sigma_L\ \sigma_R$).[49,51,52]

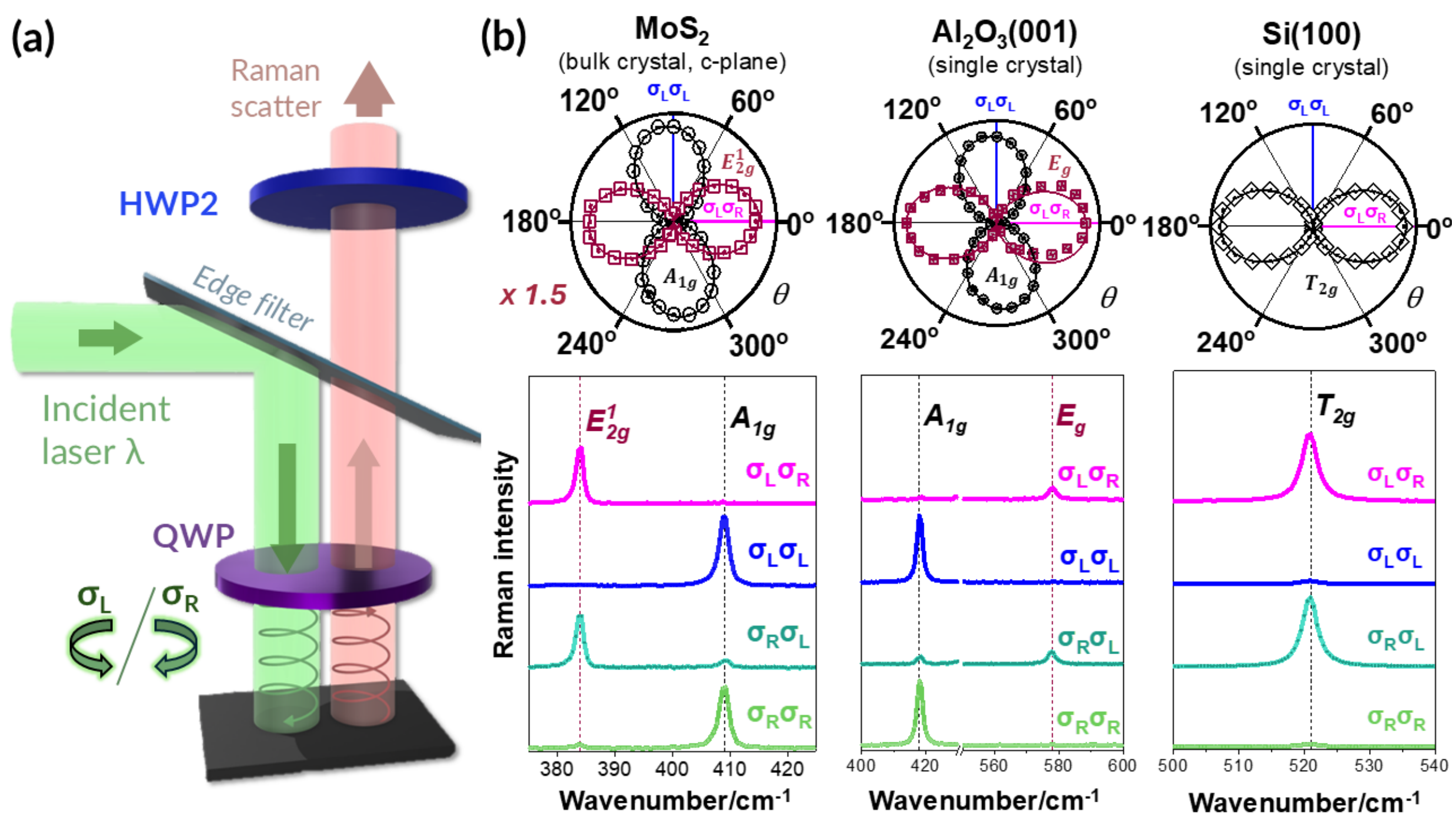

**Figure 6.** **(a)** Cartoon schematic of the optical path for helicity-resolved Raman using the combination of a quarter-wave (QWP) and the HWP2, which is used for the selection of the helicity. The mirror between the edge filter and HWP2 is not show here. **(b)** Polar plots of the integrated area of circular polarization sensitive phonon modes for $MoS_2$ ($E^1_{2g}, A_{1g}$), $Al_2O_3$ ($A_{1g},\ E_g$), and Si(100) ($T_{2g}$) as a function of $\theta$. The excitation polarization ($\lambda$ = 532 nm) is set to the circularly left ($L$ or $\sigma_L$) as it interacts with each sample. Below each polar plot is the Raman responses of the modes at four distinct helicity configurations: *LR* ($\sigma_L\sigma_R$), *LL* ($\sigma_L\sigma_L$), *RR* ($\sigma_R\sigma_R$), *and RL* ($\sigma_R\sigma_L$). Note: Sizeable differences in scale.

It is important to recognize that circular polarization arises when two orthogonal components of the electric field have equal amplitudes but differ in phase by π/2 (a quarter of a wavelength). This phase difference causes the electric field vector to rotate in a circle as the wave propagates, reflecting the rotational symmetry inherent in the polarized state. A basis transformation can be applied to transform Raman tensors (typically expressed in the Cartesian basis) into the circular basis, from which the circular polarization responses of these typical Raman phonons can be derived.[53] Two typical phonons used in this study have the following helicity dependent: i) $A_g/A_{1g}$ retains a similar selection rule as in linear polarization, where it only shows up in co-rotation ("parallel") configuration; ii) $E_g/E_{2g}$ now shows the opposite selection rule to $A_g$, where it only shows up in counter-rotation ("crossed") configuration. These rules can have exceptions in resonant Raman scattering, where the resonant electronic transition can impart an effect on the circular polarization, and in chiral materials, where the chiral phonon response is an active field of research. We strongly stress that understanding the basic circular Raman response of a non-chiral material is a prerequisite to understanding the more complex case of chiral phonons.

Similar to the linear polarization protocol, we use the Raman response of the three test materials $MoS_2$, $Al_2O_3$, and Si(100) to determine the co-rotating and counter-rotating helicity configurations. Crucially, the QWP is calibrated prior to any helicity-dependent measurements to ensure accurate polarization control. While a polarimeter is used for QWP characterization, we also demonstrate that the Raman response of $MoS_2$ can serve as an independent calibration method, eliminating the need for a polarimeter (**Figure S15**).

Setting the QWP to produce circularly polarized light (in this case $\sigma_L$), we measure the Raman response of the material as a function of HWP2 angular rotation. **Figure 6b** summarizes the polar plots obtained from the integrated Lorentzian peak areas of the $\sigma_L$ polarization-sensitive phonon modes in $MoS_2$ ($A_{1g}$, $E^1_{2g}$), $Al_2O_3$ ($A_{1g}$, $E_g$), and Si(100) ($T_{2g}$). While for linear polarization, the $E^1_{2g}/E_g$ modes in $MoS_2$ and $Al_2O_3$ are independent of $\theta$ angular rotation, excitation with circularly polarized light yields a 'peanut'-shaped polar plot ($cos^2(\theta)$) with the maximum response ($\theta = 0^0$) corresponding to $RL$ or $LR$ and the minimum ($\theta = 90^0$) for $RR$ or $LL$. On the contrary, the $A_{1g}$ and $T_{2g}$ modes exhibit an opposite response compared to $E_{2g}^{1}/E_{g}$ modes.[49,52] We note that the angles, $\theta = 0^0$ and $\theta = 90^0$ corresponds to the parallel and cross linear polarization configurations of HWP2, respectively (**Figure 6a)**.

For this reason, the second method follows the flow-chart in **Figure 1b**, showing the calibrated HWP2 (described in the Section 2.3) prior to placement of the QWP. In this case the QWP is also characterized by the polarimeter to produce $R$ or $L$. In practice, with the QWP inserted, the $RR$ or $LL$ response is measured with the HWP2 at $\theta = 90^0$ (and $RL$ or $LR$ $\theta = 0^0$). This is "opposite" of the *linear* configuration for the HWP2, where the effect comes from a $\pi$-phase shift that occurs at the back-reflection on the sample.[18,39] Following the Jones formalism, if we inspect the incident laser polarization and then the Rayleigh scattering just above the QWP, the linear polarization of the Rayleigh scatter is always rotated $90^0$ relative to the incident laser linear polarization.

## 5. Conclusions

Raman polarization measurements can be affected by misalignment of the laser and optics, sample quality, and distortions from optical components. Achieving precise, angle-resolved polarization measurements requires a dependable methodology along with robust calibration materials and techniques. This study provides an in-depth exploration of the challenges and nuances involved in obtaining accurate polarization data. We introduce an optimized methodology for Angle-Resolved Raman Spectroscopy (ARRS) that directly addresses key experimental challenges, especially those stemming from polarization distortions introduced by the dichroic edge filter. Our methodology was validated through the use of 2H-$MoS_2$ and *c*-plane $Al_2O_3$ (001) crystal; more specifically, the isotropic response of their Raman active $A_{1g}$ mode was used to calibrate the half-wave plate of our polarization analyzer. We also identified and quantified the impact of the dichroic edge filter on the anisotropic polarization response of materials such as Si(100). With incident linearly polarized light, the resulting Raman scatter from Si(100) has a mixed polarization, which induces an additional relative phase shift as the scattered light is transmitted through the dichroic filter. Modeling the edge filter as a waveplate allowed us to accurately predict the experimental polarization response. From this model, we determined that this effect is unique to each dichroic filter and cannot be entirely isolated or minimized. We also explored the VBG technology, where our findings highlight the absence of any depolarization effect of these commonly used filters.

Even so, this work provides a guiding benchmark that enables future optical filter technology to isolate these effects from Raman measurements. Moreover, the adaptation of our setup to accommodate helicity-resolved Raman experiments broadens the applicability of this technique, especially for chiral phonon characterization. However, a more detailed investigation is required to understand the impacts of dichroic edge effects with circularly polarized light.

## 6. Acknowledgements and Notes

†T. Adel, and M.F. Munoz contributed equally to this work.

M.F.M. and R.T. would like to acknowledge the NIST/National Research Council Postdoctoral Research Associateship Program for funding. This work was performed with funding from the CHIPS Metrology Program, part of CHIPS for America, National Institute of Standards and Technology, U.S. Department of Commerce.

Commercial equipment, instruments, and materials are identified in this paper to specify the experimental procedure adequately. Such identification is not intended to imply recommendation or endorsement by the National Institute of Standards and Technology or the United States Government, nor is it intended to imply that the materials or equipment identified are necessarily the best available for the purpose.

The authors declare no competing financial interests.

## 8. TOC

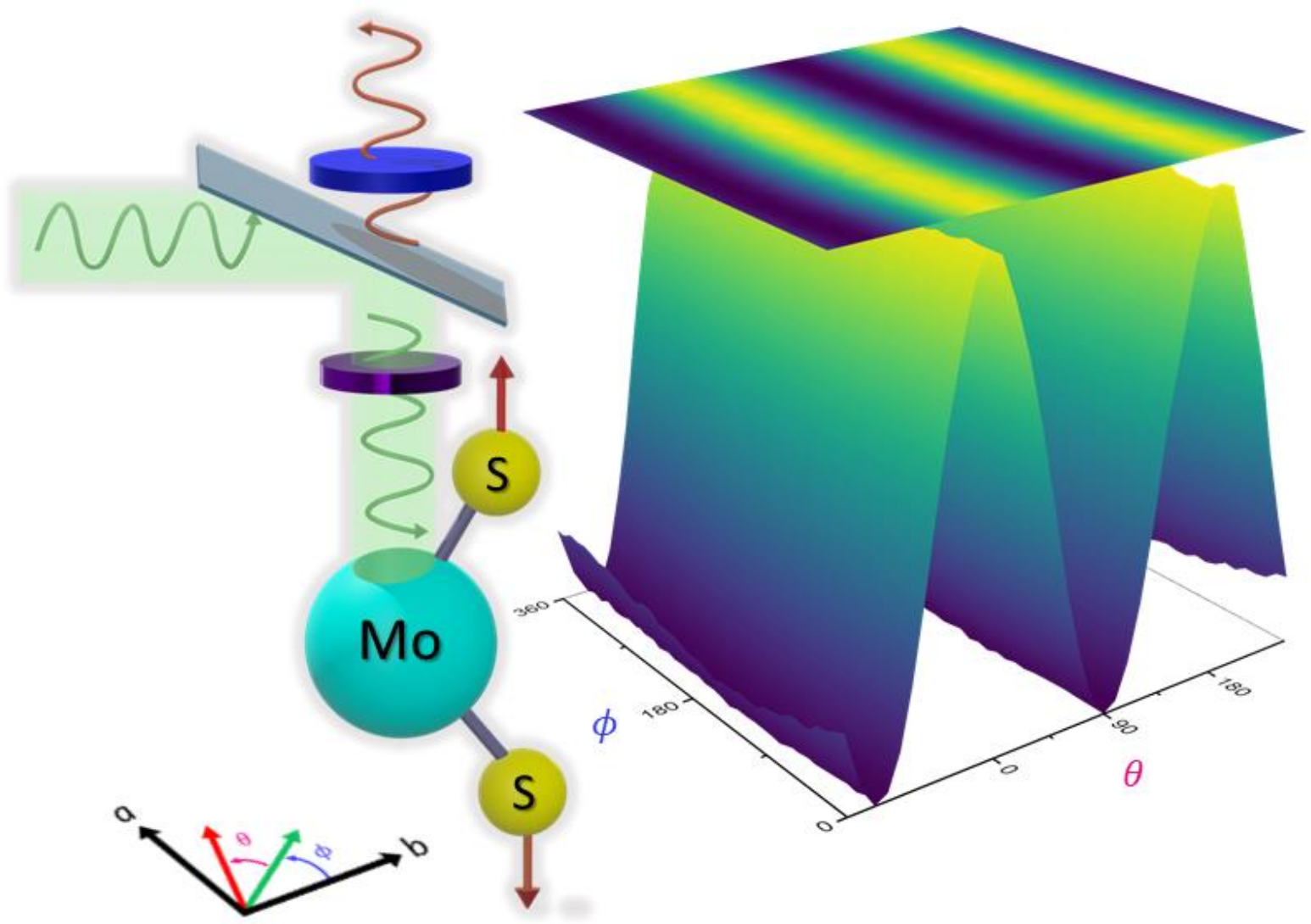

# Supplemental Information

# Accurate Angle-Resolved, Polarized Raman Methodology -Quantifying the Dichroic Edge Filter Effect

Tehseen Adel[1,2†], Maria F. Munoz[1†], Thuc T. Mai[1], Charlezetta E. Wilson-Stokes[1,3], Riccardo Torsi[1], Kevin M. Linne von Berg[4], Aurélien Thieffry[5], Jeffrey R. Simpson[1,4], and Angela R. Hight Walker[1*]

[1]Quantum Measurement Division, National Institute of Standards and Technology, Gaithersburg, MD, USA
[2]Department of Physical Sciences, University of Findlay, Findlay, OH, USA
[3]Department of Mechanical Engineering, Howard University, Washington, DC, USA
[4]Department of Physics, Astronomy and Geosciences, Towson University, Towson, MD, USA
[5]HORIBA France, Lille, Hauts-de-France, France

## S1. Placement of the Si(100) wafer

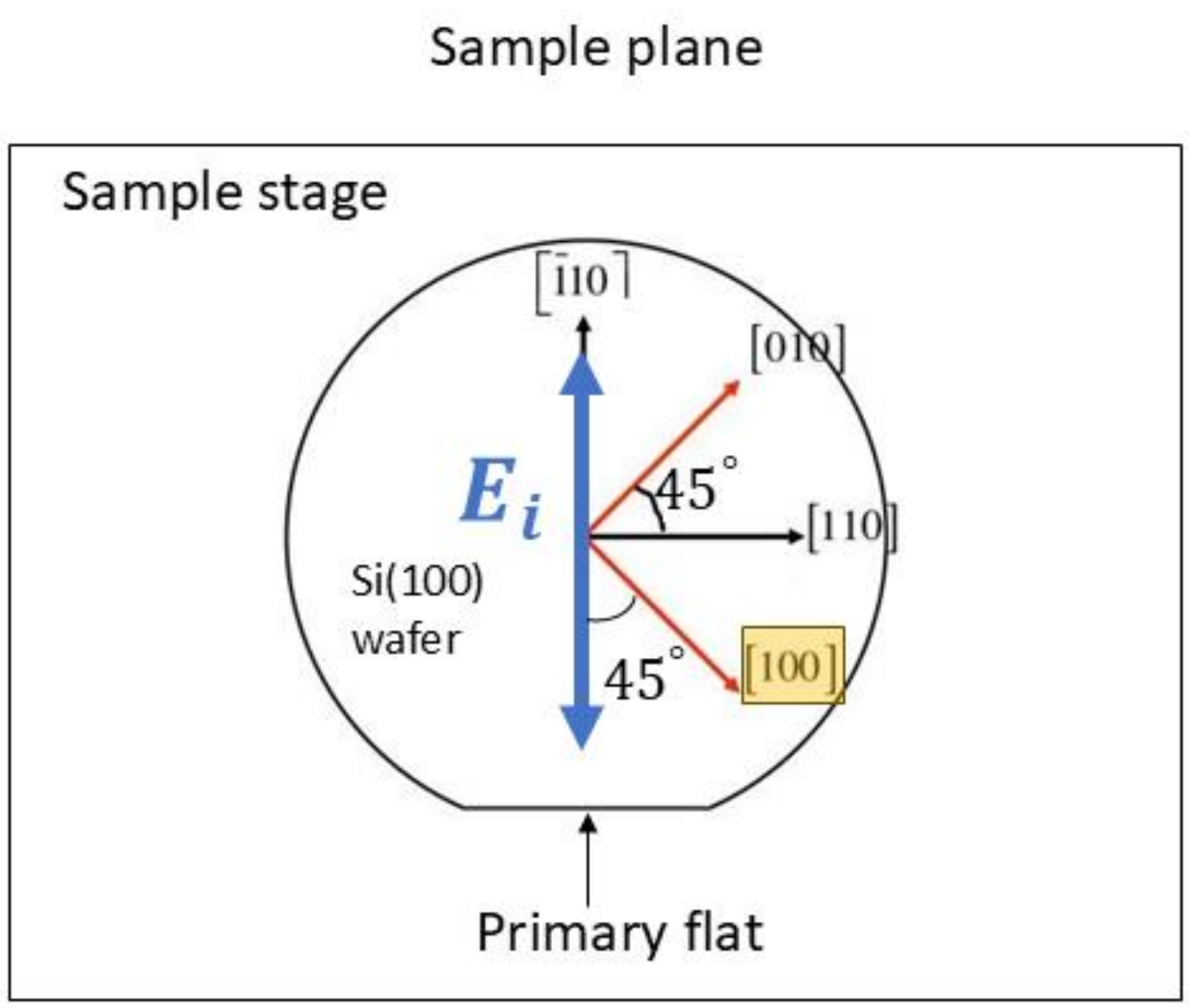

**Figure S1.** Placement of the Si(100), where the incident light ($\boldsymbol{E_i}$) is perpendicular to the flat edge "primary flat", and 45° from the plane [100]. The figure is the top view of the sample stage, where the incident light *k* vector is in the *z*-axis out of the paper.

## S2. The performance of the half-wave plates (SA-HWPs) is affected by their location

Before testing the linear polarization protocol, the super achromatic half-wave plates (SAHWPs) were characterized using a polarimeter. This was an important and necessary step in our methodology, enabling us to identify changes in polarization resulting from optics in the beam path.

As shown in **Figure S2a.**, light from the source first passes through polarizer P1, fixing the initial polarization. Before reaching the sample, the beam passes through the HWP1, which controls the direction of polarization for the incident light. For HWP characterization and to optimize optics placement, we tested two configurations (see **Figure S2a.** and **S2b.**): configuration 1, in which the SA-HWP is placed ***after*** the edge filter, and configuration 2, where the SA-HWP is placed ***between P1 and the edge filter***. The polarimeter is placed at the sample position to measure the measurands for both configurations. Rotating the SA-HWP by 5° steps, we measure and report the ellipticity values and light polarization ($\beta$) for each configuration here in **Figure S2**. Using this information, the SA-HWP is then placed and configured such that the incident light zero-point has vertical (*V*) polarization with respect to the polarimeter frame, i.e., the polarization value $\beta = 90°$, measured by the polarimeter as shown in **Figure S2a,** leftmost panel.

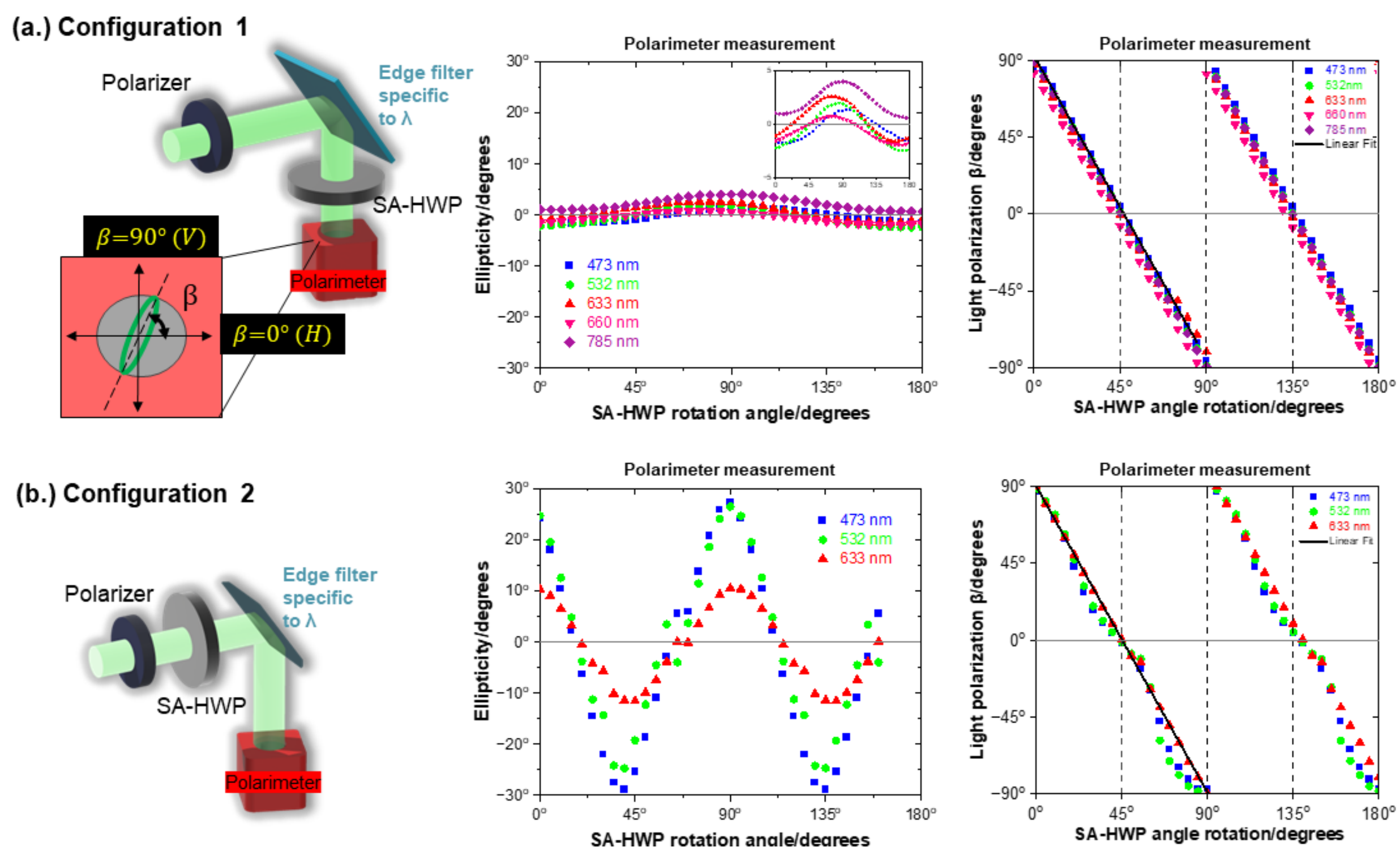

**Figure S2.** Two linear polarization configurations of the laser input and their polarimeter characterization (ellipticity, angle of incoming light polarization) of the super achromatic half-waveplate (SA-HWP): (a.) The SA-HWP is placed after the edge filter that shows a ±4° change in ellipticity. (b.) With the SA-HWP placed before the edge filter, the change in ellipticity is nearly ±30° for the shorter wavelength lasers. Such large values adversely affect polarization measurements. Solid lines show a linear fitting of the data.

This characterization highlights the effect of HWP placement on the ellipticity of the incident light, which will affect how the light interacts with the sample. The configurations tested here have been presented and discussed in the literature.[1] The placement of SA-HWP in configuration 1 results in the incident light having our desired or chosen polarization $\phi$. However, when the back-scattered light passes through the SA-HWP, its polarization is restored to the initial polarization, i.e., *V* polarization. **Figure S2a.** shows the schematic of configuration 1 (left panel), as well as the ellipticity (middle plot) and measured polarization (right plot) for several wavelengths as the SA-HWP is rotated from 0 degrees to 180 degrees. For wavelengths within the visible (473 nm, 532 nm, 633 nm, 660 nm, 785 nm), the ellipticity introduced by the edge filter reflected beam remains within $\pm 4°$. The maximum ellipticity observed is 4.25° for 785 nm light transmitted through the SA-HWP rotated to 90°. Each wavelength has maximum ellipticity at different positions of SA-HWP, however for 633 nm, 532 nm and 473 nm, the ellipticity is minimized to 0° at 2 positions. We observed that ellipticity variation at these small values did not affect polarized Raman measurements for the samples studied in this work. The middle plot of **Figure S2a.** shows that all wavelengths follow the expected linear behavior with slope of 2 obtained from fitting the experimental data. There is no observable deviation from this linear dependence for all tested wavelengths.

**Figure S2b.** shows configuration 2 where the SA-HWP is placed ***before*** the edge filter. Note that, for this configuration, only 3 wavelengths (473 nm, 532 nm, 633nm) were measured. This configuration provides control of the incident light without altering the scattered light. The range of ellipticity variation for this configuration increases an order of magnitude to $\pm 30°$, highlighting that, in this configuration, polarized light experiences an induced ellipticity from its interaction with the dichroic edge filter. Hence, the resulting reflecting incident beam will have a

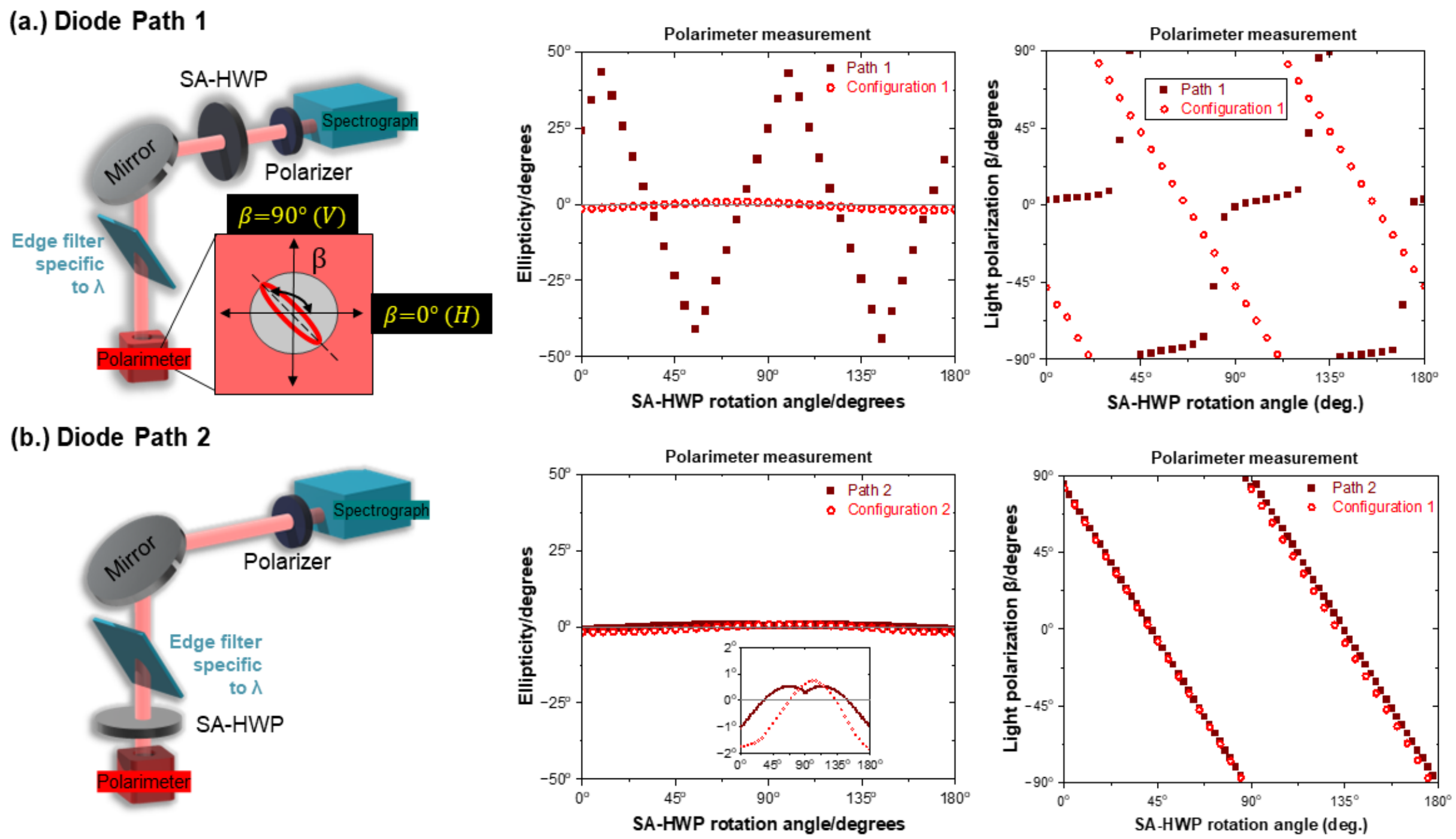

**Figure S3.** Polarization configurations showcasing two different locations of the super achromatic half-waveplate (SA-HWP) in the Raman scatter path. Using a polarimeter, the ellipticity and degree of light polarization is characterized with a 660 nm diode laser (housed within the spectrograph). The results are compared to Configuration 1 in Figure S1. (a.) The change in ellipticity for the diode Path 1 (±45°) is larger relative to Configuration 1 (±4°). (b.) Similar to Configuration 1, the SA-HWP placed between the edge filter and polarimeter has a similar change in ellipticity (±2°) for that specific wavelength.

different polarization than expected. When the SA-HWP is rotated close to 45°, 90°, and 135°, the incident light is no longer linear but elliptical, thus affecting polarized Raman measurements. The primary effect from the ellipticity variation in this configuration is shown in the middle plot of **Figure S2b**., where one can see deviations from the expected linearity observed in configuration 1 for each wavelength tested.

For additional insights of the edge filter effect on the ellipticity or polarization of the light, we explored the scattered light path using the instrument's internal diode laser as shown in **Figure S3a.** and **S3b.** In this configuration, the second polarizer (P2) acts similarly to P1 in the excitation pathway, fixing the polarization to vertical ($\beta = 90°$) as measured by the polarimeter at the sample position, without the SA-HWP in the path. Following the schematic for this configuration (**Figure S3a.**), the light has its polarization fixed by P2, which can then be modified as the SA-HWP is rotated. After the polarization is set by the SA-HWP, contrary to the excitation path configurations, the light is transmitted through the edge filter. **Figure S3a.** (middle plot) shows that ellipticity is measured up to 45 degrees, generating right and left circularly polarized light. The range of ellipticity (±45°) is significantly higher than the ellipticity observed from configuration 2 of the excitation path. This behavior suggests the combination of the SA-HWP with the edge filter causes the light to become polarized as if it has been transmitted through a quarter waveplate. This effect is outlined in the main body of the paper **Section 3.2**. There are periodic SA-HWP positions where the ellipticity is zero and the light is purely linear. However, materials with phonons sensitive to circularly polarized light will exhibit additional effects in the Raman response when this optical configuration is used. Thus, it is critical to select the appropriate configuration to perform angle resolved Raman spectroscopy measurements. To avoid distortion in the polarization in Raman measurements, according to our results we recommend following **Configuration 1**.

In **Figure S3a**. (right plot), rotating the SA-HWP causes the direction of the light polarization to exhibit unusual behavior when compared to the expected linear response observed in configuration 1 of the excitation path (red circles). At the initial position of the SA-HWP, the light polarization is horizontal ($H$; *i.e.,* $\beta = 0°$) and remains so until the SA-HWP is rotated 45° from its initial position, at which point the polarization becomes vertical ($\beta = \pm 90°$). The trend continues, with light polarization jumping from $H$ to $V$ every 45°. For the same path we repositioned the SA-HWP ***after*** the edge filter (**Figure S3b.**) and repeated the measurements collected for diode path 1. This path is analogous to configuration 1 for excitation path in the sense that the HWP is placed ***after*** the edge filter. However, in the excitation path we examined edge filter effects on reflection, while in the diode path we examine transmission effects. Placing the HWP after the edge filter passivates the depolarization effect observed in diode path 1. The polarization response is restored to the expected linear tendency as the SA-HWP is rotated (right plot) and the ellipticity variation is reduced to ±2° for all wavelengths.

Based on the characterization of our optics in different configurations along both the excitation and scattering paths, we determined that placing the first SA-HWP ***after*** the edge filter would produce the best results. By choosing configuration 1, depolarization effects due to excitation light reflection are minimized. However, by using this configuration, transmission of back scatter through the first SA-HWP and then through the edge filter, cannot be avoided. Thus, in our model we cannot neglect transmission effects on the Raman response during the collection of 2D polarization maps.

**S3. Alternative protocol for polarizer P1**

Here we present how to align the polarizer P1 without a polarimeter to set the input laser polarization along the *x* axis in the sample plane as is indicated in **Figure 1(a.).** Note that without a polarimeter the procedure requires a wired grid polarizer on glass substrate (Thorlabs, WP12L-VIS), as reference. Optically aligning the axis of the polarizers in a direction perpendicular to the light propagation is determined by measuring the optical power transmitted through the respective optic. The power output through P1 ($P1_V$) is proportional to the linearly polarized input of the laser ($P1_0$) and the cosine-square of the angle $\rho$ between the vertical axis and the linearly polarized input: $P1_V = P1_0 \cos^2(\rho)$.

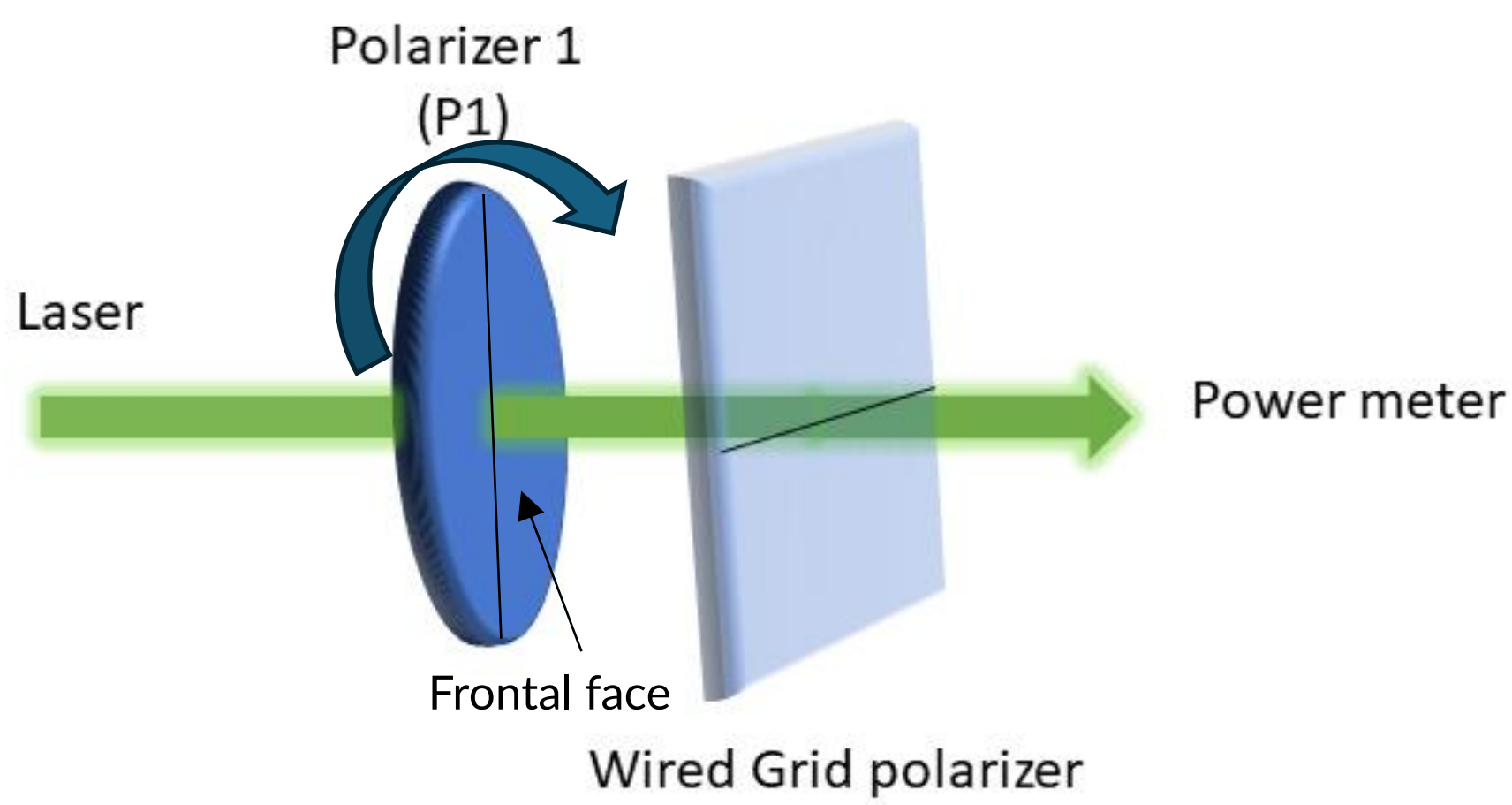

**Figure S4**. Cartoon schematic of the P1, the wired grid polarizer with the polarization axis in cross with initial polarization set by P1 along the *x*-axis, and power meter sensor.

Once P1 is placed, the wired grid polarizer film is set in front of P1 as shown in **Figure S4**. The polarizer film has a reference mark for its polarization axis which makes easy to place the axis crossing the desired polarization fixed by P1(see **Figure S4)** along the *x* axis. i.e. the polarization film axis is aligned along the *y* axis in the sample plane The polarization film is acting as our reference, and the power meter is used to measure the output power as P1 is rotated. The minimum power (close to zero) will correspond to the position of P1 that fixes the laser polarization along the *x* axis. Once the polarizer film is removed, P1 is correctly aligned along the *x* axis when the measured optical power has the same magnitude as it passes through both faces (frontal and back) of the polarizer. The polarizer optic must be flipped from one face to the other face as the power is being measured.

**S4. Calibration of HWP2 using the response of $A_{1g}$ mode of 2H-$MoS_2$.**

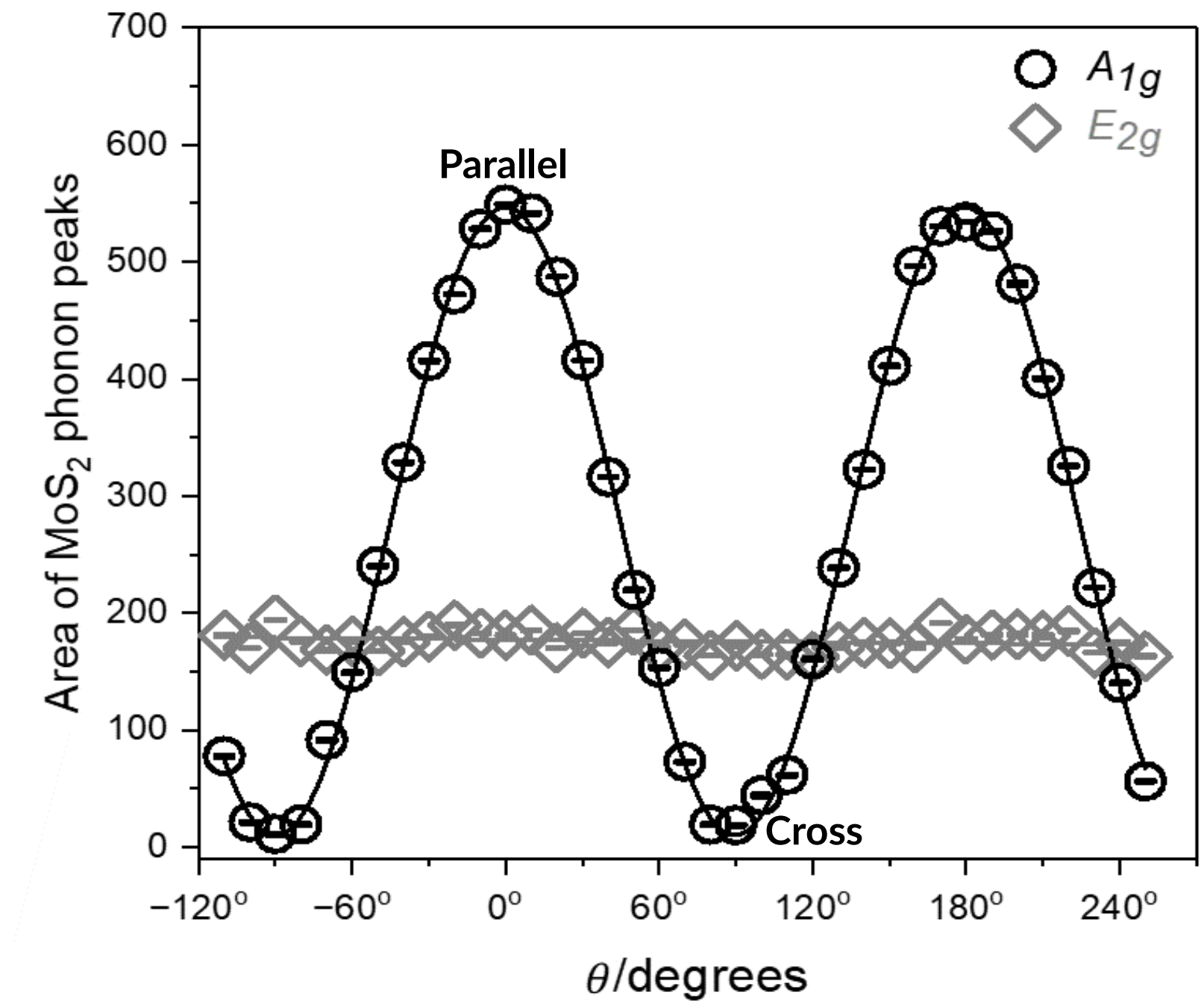

**Figure S5.** Plot of the fitted area of the $MoS_2$ Raman active modes and their polarization response as a function of a SA-HWP2 angle in 5-degree increments (error bars show the standard error from Lorentzian peak area fitting). The maximum area of $A_{1g}$ corresponds to the parallel polarization, which is the $\theta = 0°$ and 180° positions. The cross polarization is at the point where the $A_{1g}$ minimum is at $\theta = 90°$.

## S5. Alternative method for half waveplate HWP1 calibration

This procedure can be used to calibrate the half waveplate in the HWP1 motorized mount (**Figure S6a**) without using a polarimeter. Note that this procedure requires an additional polarizer (P0) and a power meter. After calibrating polarizer 1 (P1) as described in **Section 2.3**, or following the **Section S3**, the next step is to place P0 and the power meter sensor under the objective turret such that the beam passes through P0 and is incident on the sensor (**Figure S5b)**. With the P1 fixed along the x axis, rotate P0 until the optical power measures near zero; this means that P0 polarization is perpendicular to the polarized laser output through P1. With P1 and P0 in this "cross" configuration, insert HWP-1 between the edge filter and P0. Rotate HWP1 until the power is once again minimized. This corresponds to the fast axis of HWP1 being collinearly aligned with P1, i.e., the polarization is aligned along the x axis. This is the $\phi = 0°$ set point.

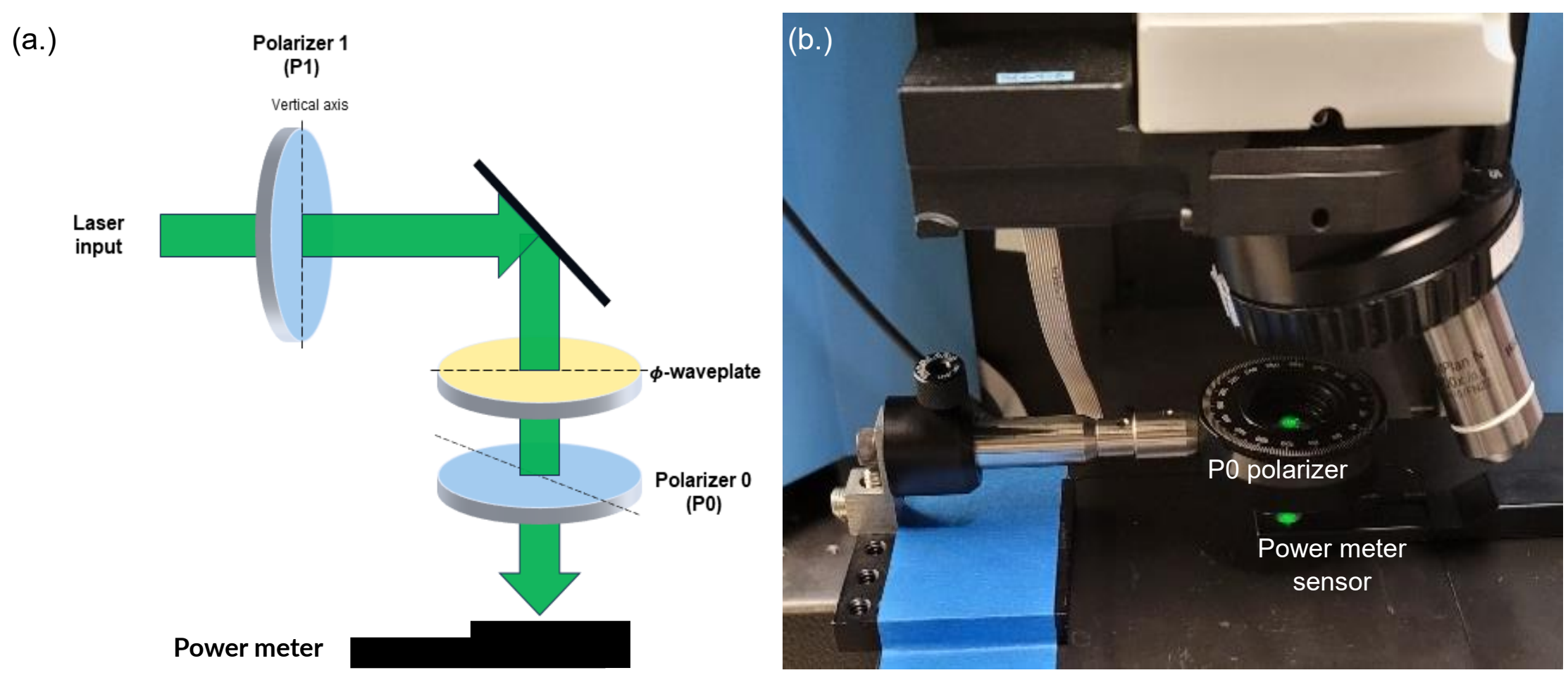

**Figure S6**. (a.) Cartoon schematic of the P1 and P0 polarizers, $\phi$-waveplate, and power meter sensor. (b.) The waveplate is "sandwiched" between P1 and P0.

## S6. Angle resolved Raman (ARR) response of c-cut $Al_2O_3$ (sapphire)

In this section, we demonstrate that sapphire is an option for calibration purposes if 2H-$MoS_2$ is not available. We select c-cut $Al_2O_3$ and perform polarized Raman measurements, examining the response of the $A_{1g}$ and $E_g$ phonons (as done for $MoS_2$) using the same wavelengths. $Al_2O_3$ is more resistant to environmental degradation or laser-induced damage compared to $MoS_2$. However, as **Figure S7a.** shows, the signal to noise (S/N) ratio for the $E_g$ mode in $Al_2O_3$ is low when compared to the $E^1_{2g}$ in $MoS_2$. Thus, the acquisition time of the Raman spectra is longer and requires more power for all wavelengths utilized. The table (**Figure S7b.**) shows the polarization response of $A_{1g}$ mode for $Al_2O_3$. The top row of polar plots show the response under $\theta$ rotation, displaying the expected "peanut-shape" for all the tested wavelengths. We note that we have better agreement compared with the theoretical prediction for 633 nm than $MoS_2$ due to the resonance effect in $MoS_2$. If the Raman system only has 633 nm excitation laser, we recommend using this crystal for calibrating the HWP2. For other colors, its response is consistent with $MoS_2$ results, allowing appropriate calibration of the HWP2 using the $A_{1g}$ mode of $Al_2O_3$.

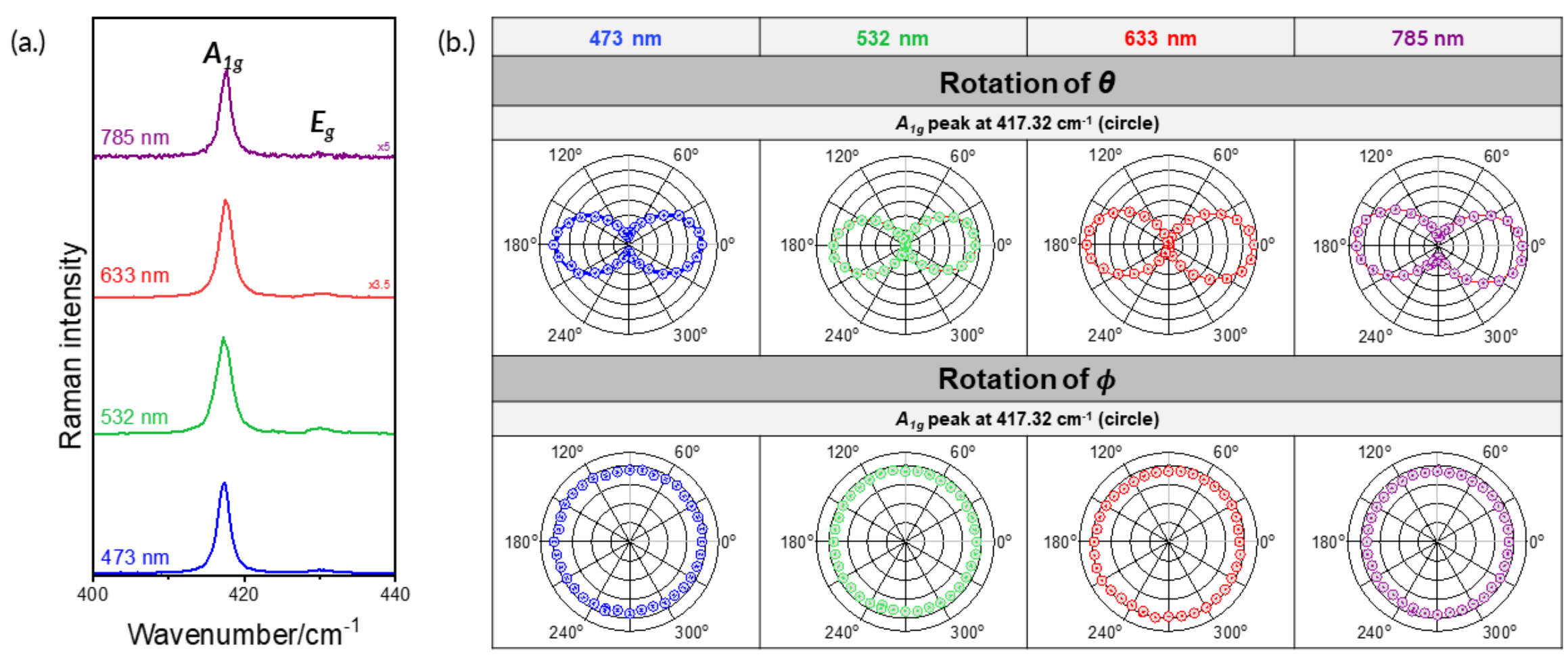

**Figure S7.** (a.) Raman response from, c-cut $Al_2O_3$ (sapphire) with four distinct excitation wavelengths showing the out-of-plane $A_{1g}$ (415 $cm^{-1}$) vibrational modes collected with a 50x objective. (b.) Polar plots of the $A_{1g}$ (open circles) integrated area (Lorentzian fitting) are plotted as a function of the two different half-wave plate rotations, revealing their complementary polarization characteristics.

**S7. ARRS response of bulk $MoS_2$ using a 100× objective.**

**Figure S8** shows the Raman response from bulk $MoS_2$, as explained in detail in the main body, but using a 100× objective. The S/N of these spectra is comparable to that observed with the 50× objective. The analysis of data to get the polar plots for $\theta$ and $\phi$ rotations, were done in the same way as when using the 50× objective. The obtained polar plots shown in **Figure S8b**, for excitations 473 nm, 633 nm and 532 nm are very similar to **Figure 2**. In the main body of the paper. However, for 785 nm, using the 100× objective we notice a slight asymmetry in the circle shape for $E_{2g}$ mode (far right column of **Figure S8b.**), and the $A_{1g}$ mode response not going to zero at the minimum at ($\theta = 90°$) showing a "open peanut-shape" not closing**.** We attribute these effects to the NA of the objective; smaller (0.55 NA for 50×) preserves the polarization better than larger NA (0.9 N.A.). Based on that finding, we performed all polarization scans using a 50× objective in the main body of the paper.

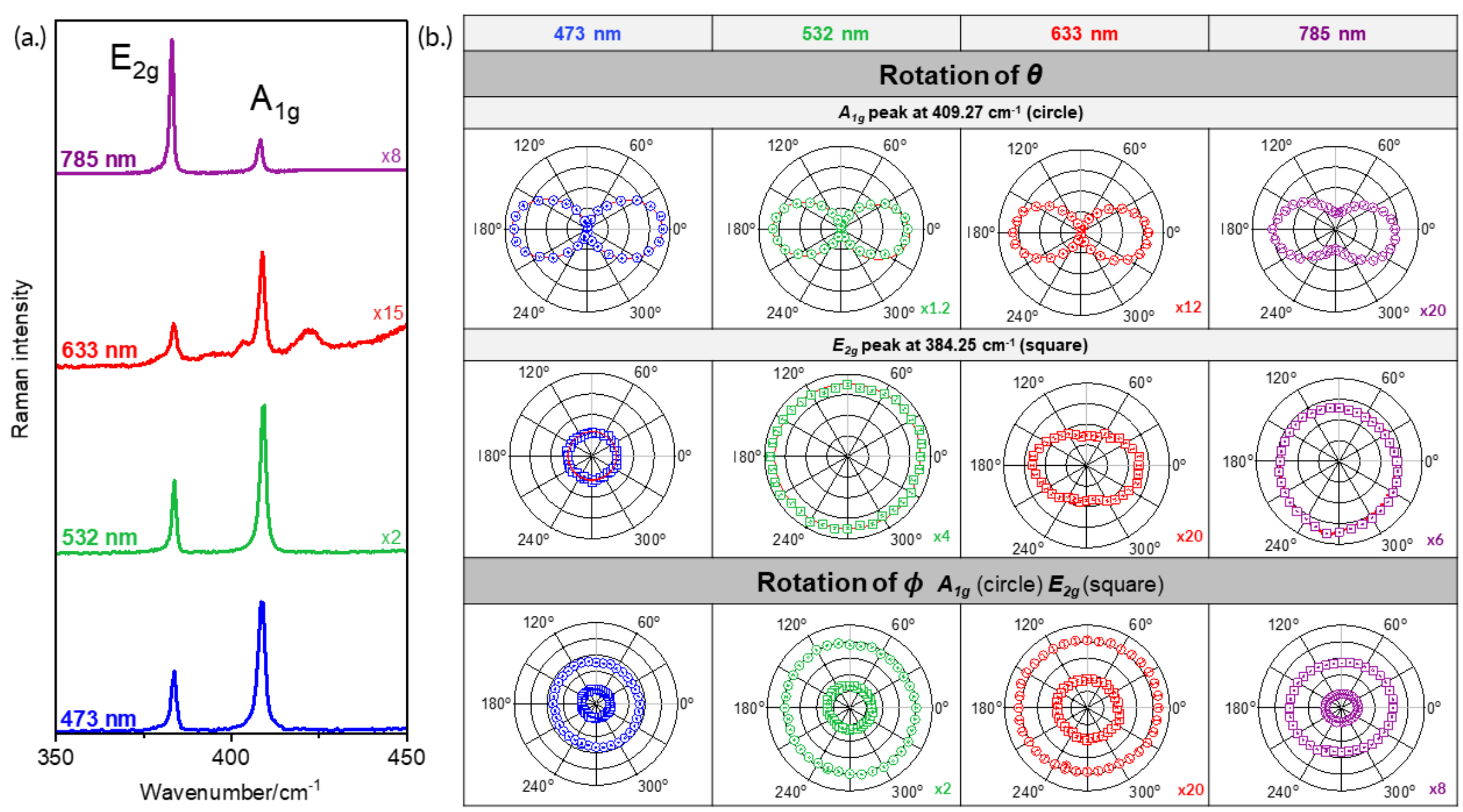

**Figure S8.** (a.) Raman response from bulk $MoS_2$ at four distinct excitation wavelengths showing $E_{2g}$ (384.25 $cm^{-1}$) and $A_{1g}$ (409.27 $cm^{-1}$) vibrational modes collected with a 100x objective. (b.) Polar plots of the integrated area (Lorentzian fitting) under the $A_{1g}$ (open circles) and $E_{2g}$ (open squares) peaks are plotted as a function of the two different half-wave plate rotations, revealing their complementary polarization characteristics. Note: Sizeable differences in scale.

## S8. Numerical aperture effect on ARRS 2D maps: $MoS_2$, Si (111), and Si (100)

To investigate the effect of numerical aperture (NA) on polarized Raman data, we show 2D polarization maps for three materials [2H-$MoS_2$, Si(111), and Si (100)] using four different objectives with magnification (and NA) as follows: 10X (0.25), 50X long-working distance (0.5), 50X (0.75), and 100X (0.9). **Figure S9** shows the resulting 2D polarization maps using a 532 nm laser excitation. Integration times were adjusted (increased for low NA objectives) to maintain a similar signal-to-noise ratio (SNR). The measured 2D polarization maps closely match the predicted responses anticipated from the Raman tensors for the $A_{1g}$ and $T_{2g}$ modes, respectively,

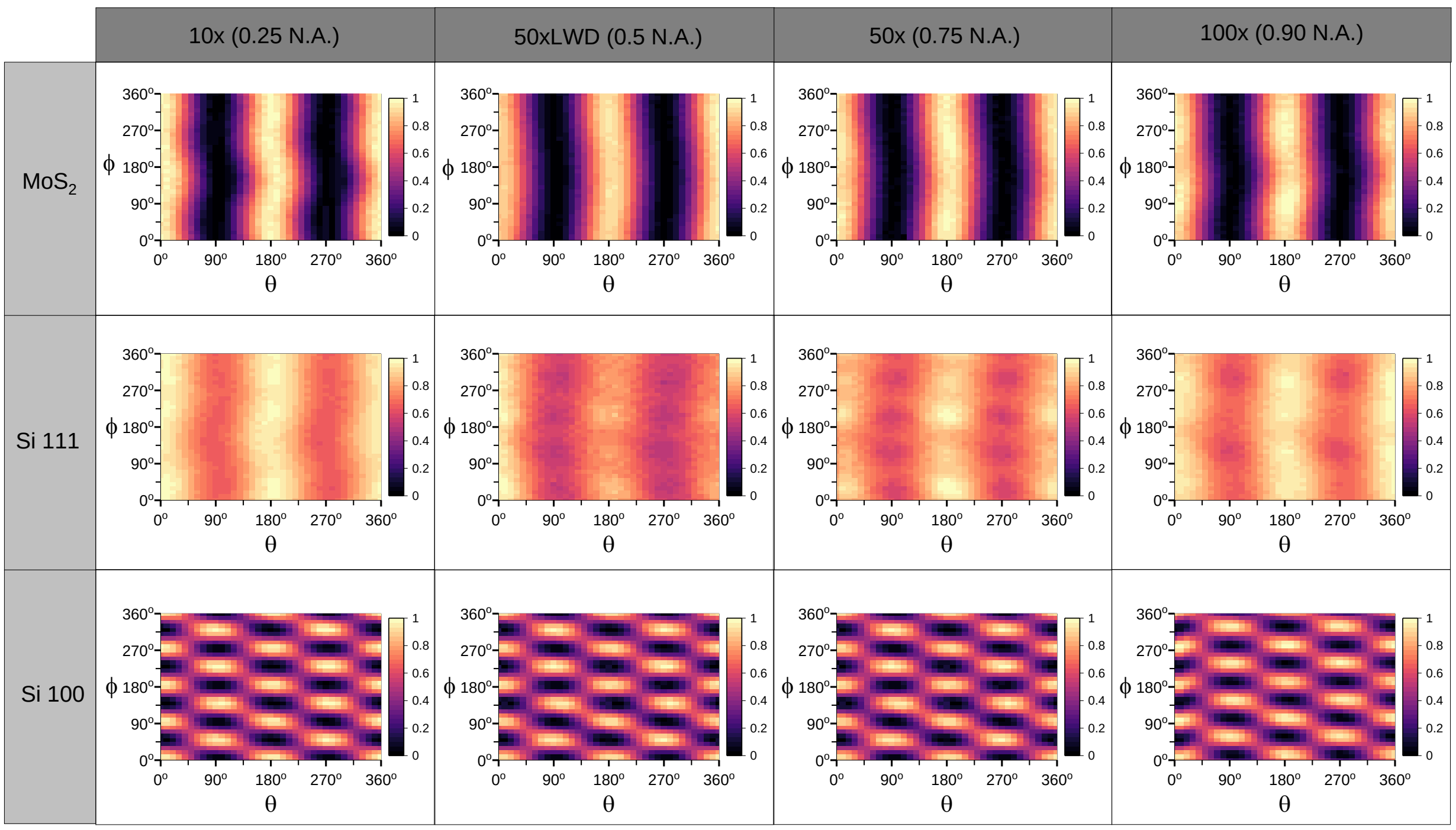

**Figure S9.** 2D polarization maps of various samples with isotropic response collected with different N.A. objectives: $MoS_2$ – $A_{1g}$ peak area, Si(111) and Si(100) – $T_{2g}$ peak area.

for $MoS_2$ and Si. We observe no discernible difference in Raman maps for various NA objectives.

In order to quantify the agreement between experimental 2D polarization maps and the predicted model, we calculate the residuals (estimated variance)

$$\text{Residual} = \frac{1}{N}\sum_i^N \left[I_{\text{exp},i}(\theta,\varphi) - I_{\text{mod},i}(\theta,\varphi)\right]^2 \qquad \text{(S1)}$$

where N is the total number of spectra in a given 2D map (37x37 =1369) and $I_{exp,i}(\phi,\theta)$ and $I_{mod,i}(\phi,\theta)$ are the integrated peak intensities for the experimental and model spectra, respectively. Integrated peak intensities are normalized to the maximum value. Percent residuals for the samples and objectives shown in **Fig. S9** are tabulated in **Table S1**. The largest residual value is 0.0040, corresponding to 0.40%, indicating significant agreement between the experiment and the model. There is no clear monotonic behavior with objective NA, consistent with the fact that the field distribution in typical backscattering geometries remains dominated by in-plane polarization even at high NA. Subtle offsets in the vertical band shapes shown in **Fig. S9** are attributed to instrumentation anomalies associated with motorized movement of the half-wave plates.

| Objective N.A. | 10x (0.25) (%) | 50xLWD (0.50) (%) | 50x (0.75) (%) | 100x (0.90) (%) |
|---|---|---|---|---|
| $MoS_2$ | 0.37 | 0.31 | 0.40 | 0.40 |
| Si(111) | 0.06 | 0.29 | 0.13 | 0.08 |
| Si(100) | 0.14 | 0.23 | 0.15 | 0.25 |
| Si(100)-Bragg | | | | 0.07 |

**Table S1.** Percent residuals comparing 2D map data to polarization model for four different objectives in order of increasing magnification (respective numerical apertures shown in brackets) and three different material systems: $MoS_2$, Si (111), and Si (100). Si (100) using Bragg filters (last row) with a 100X objective is shown for comparison.

Despite the limited evidence for variation in either the 2D maps of **Fig. S9** or the residuals in **Table S1** across different objectives, we consider whether a higher NA might have an effect by introducing a component of the electric field normal to the sample plane. In a typical 180° backscattering configuration, the incident and scattered light propagate parallel to the sample unit normal, taken to be the z-axis here. Such normal light consists of an electric field polarized parallel to the sample plane (xy-plane) only. Conversely, non-normal light contains a component of the E-field polarized along the z-axis, $E_{z\text{-}pol} = \{0, 0, E_0\}$. To consider the effect of such z-polarized light on the Raman response, we include this in Eq. (1) along with the appropriate Raman tensors Eqs (2) and (4). Note: there is no z-polarized contribution to the in-plane Raman response from symmetric modes, which lack off-diagonal z-components, i.e., $R_{xz}$, $R_{yz} = 0$. For example, the diagonal $A_{1g}$ mode of $MoS_2$ [see **Eq. (2)**] contains only zero off-diagonal elements; therefore, incident light polarized along z does not couple to the in-plane scattered response for these samples. For materials with a Raman tensor having off-diagonal z-components, we consider a simple empirical model including z-polarized light as the following

$$I_{\text{total}}(\theta,\varphi) = \alpha\, I_z(\theta,\varphi) + (1-\alpha)\, I_{\text{n-plane}}(\theta,\varphi) \qquad \text{(S2)}$$

where $\alpha$ is a fit parameter indicating the amount of the $z$-polarized light. More sophisticated theoretical calculations of the E-field components for light focused to a diffraction-limited spot are given by Debye[2] and Wolf [3]. When we include our empirical model with Si(100), we find only a small contribution from the z-polarized light, with a value of $< 0.02$ at the highest NA (0.9), and minimal change in the residual. Therefore, we conclude that there is no significant contribution from z-polarized light, nor a significant dependence of numerical aperture, to our 2D polarization maps.

### S9. ARRS response 2D maps of Si cuts: Si (111), and Si(110)

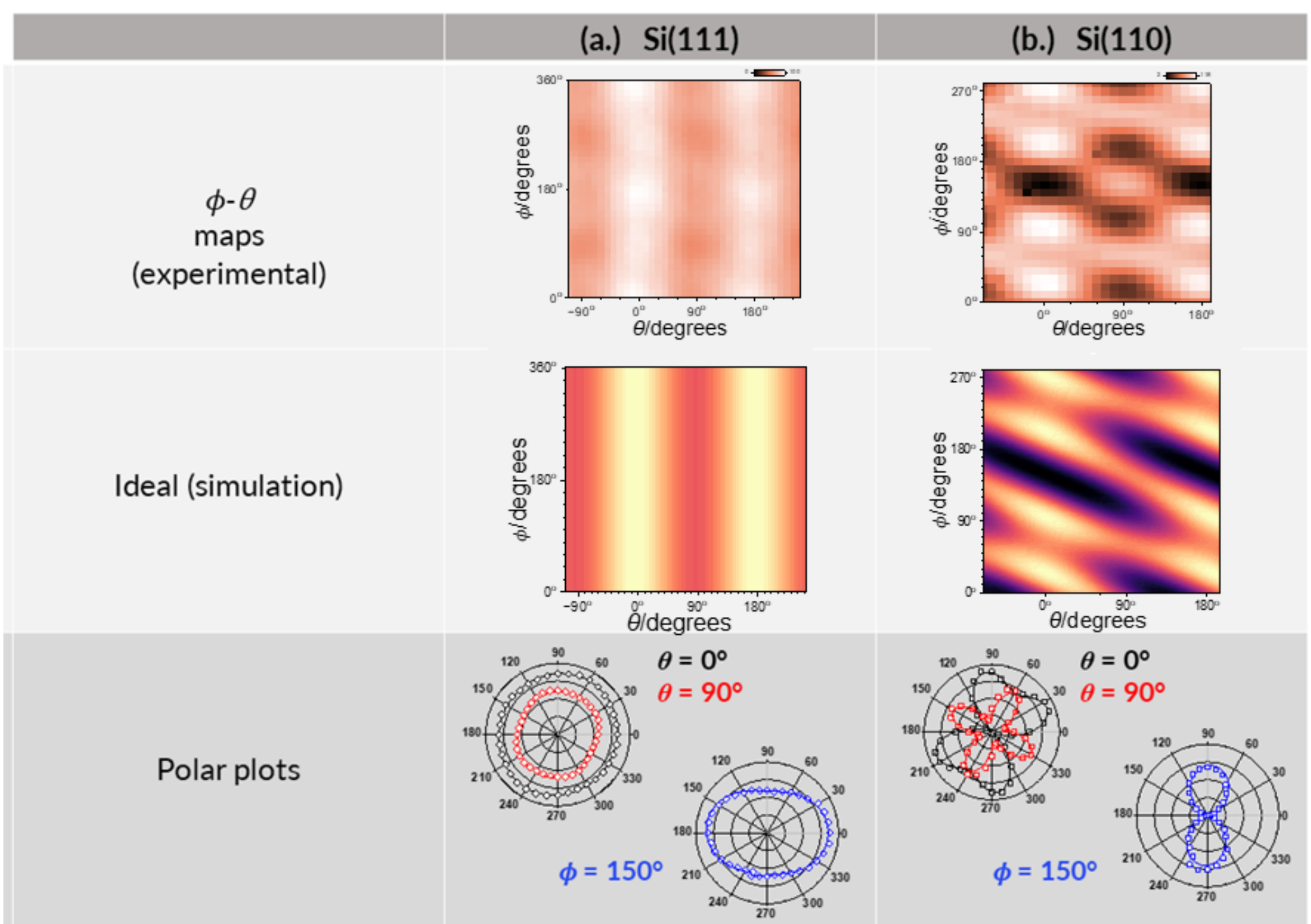

**Figure S10.** 2D polarization maps of the $T_{2g}$ peak area showing the different cuts of silicon compared to their respective calculated Raman $T_{2g}$ tensor: (a.) Si(111), and (b.) Si(110). Polar plots at θ = 0° and 90° (circles) and $\phi$ = 180° (diamonds) are shown below the 2D maps. These plots highlight the clear contrast between the Si(111) with isotropic response, and the cuts with anisotropic response of Si(100) (main body of the paper) and Si(110).

Using Jones Matrix formalism, we simulate the ideal response for the Si cuts: isotropic response (111), and anisotropic response (110), and generate the 2D polarization maps to compare our experimental 2D maps (shown in **Figure S10**). The $T_{2g}$ peak response of Si (111) is close to the ideal response, as the other materials with isotropic response discussed previously in the main text of this work. In the case of the cut Si (110) with anisotropic response (**Figure S10b**.), we see that there is an edge filter effect as was observed in the cut Si (100) (discussed in the main body of the paper), hence our case differs from the ideal case. However, we are showing that for specific configuration (parallel or cross), the polar plots agree with the ideal case for $\phi$ rotation. By fixing $\phi$ and rotating $\theta$ its more complicated since the ideal case is very complex, and the response varies depending on the incident light. We are showing that for an arbitrarily chosen slice at $\phi = 150°$ the experiment shows the expected response. This data is added for reference and is not analyzed further in this work.

## S10. Simulation of the theoretical model accounting for the edge filter effect

### A. Jones Matrix representation of a half-wave plate

The Jones matrix representation[3] of a half-wave plate with the fast axis rotated CW from the *x*-axis by angle, and used in our model, is given by

$$M_{HWP}(\alpha) = \begin{bmatrix} \cos(2\alpha) & \sin(2\alpha) \\ \sin(2\alpha) & -\cos(2\alpha) \end{bmatrix}, \tag{S3}$$

where $\alpha$ represents the physical rotation angle of the HWP. In terms of the polarization of the scattered and incident light ($\theta$ and $\phi$, respectively), $\alpha = \frac{\theta}{2}$ or $\frac{\Phi}{2}$. The sign of the angle depends on the rotation direction of the half-wave plate (CCW positive or CW negative observed from the detector).

**B. The three regions of $\delta$ delta phase**

The expression for the $T_{2g}$ phonon polarization response from a Si(100) wafer accounting for the edge filter effect, is

$$I_{T_{2g}}(\theta,\phi) \propto \left|e^{i\delta} t_p \sin(\theta)\cos(2\phi) + t_s \cos(\theta)\sin(2\phi)\right|^2, \quad \text{(S4)}$$

Where the value (and sign) of $\delta$ affects the direction of the modulated diagonal bands, where $\delta = \pi/2$ indicates the value where the bands are very symmetric, i.e., where there is no apparent slope. The bands slope towards the positive direction, where $\delta > \pi/2$ , and negative for $\delta < \pi/2$ .

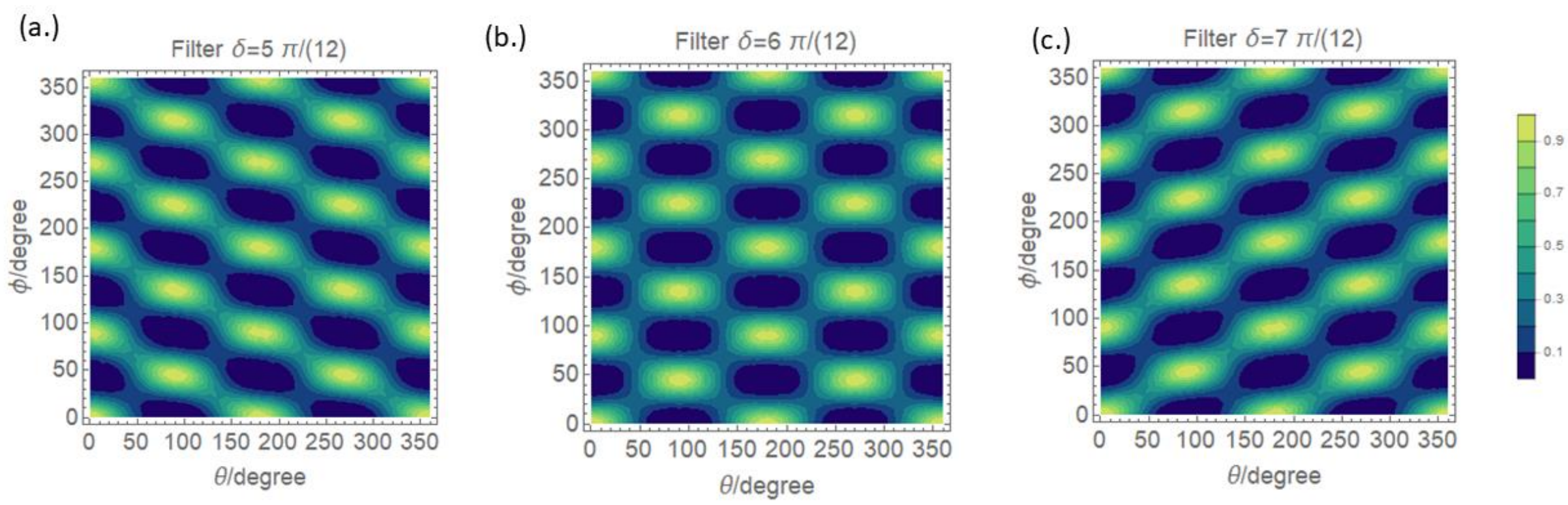

**Figure S11.** Simulations of the expression in **Eq. (8),** for different values of (a) $\delta = \frac{5\pi}{12}$ , (b) $\delta = \frac{6\,\pi}{12} = \frac{\pi}{2}$ , and (c) $\delta = \frac{7\pi}{12}$ . Showing the change of the slope depending on the value of $\delta$ phase.

## S11. Experimental data and fitting 2D polarization maps for the $T_{2g}$ Raman response of Si (100) at different excitation and edge filter wavelengths

The 2D polarization maps from the fitting of the experimental data collected at 473 nm, 633 nm, and 785 nm excitation lasers are shown in **Figure S12**, using the proposed model. From the fitting we extract the delta phase difference ($\delta$) for each color, and, as expected, each edge filter has a different value.

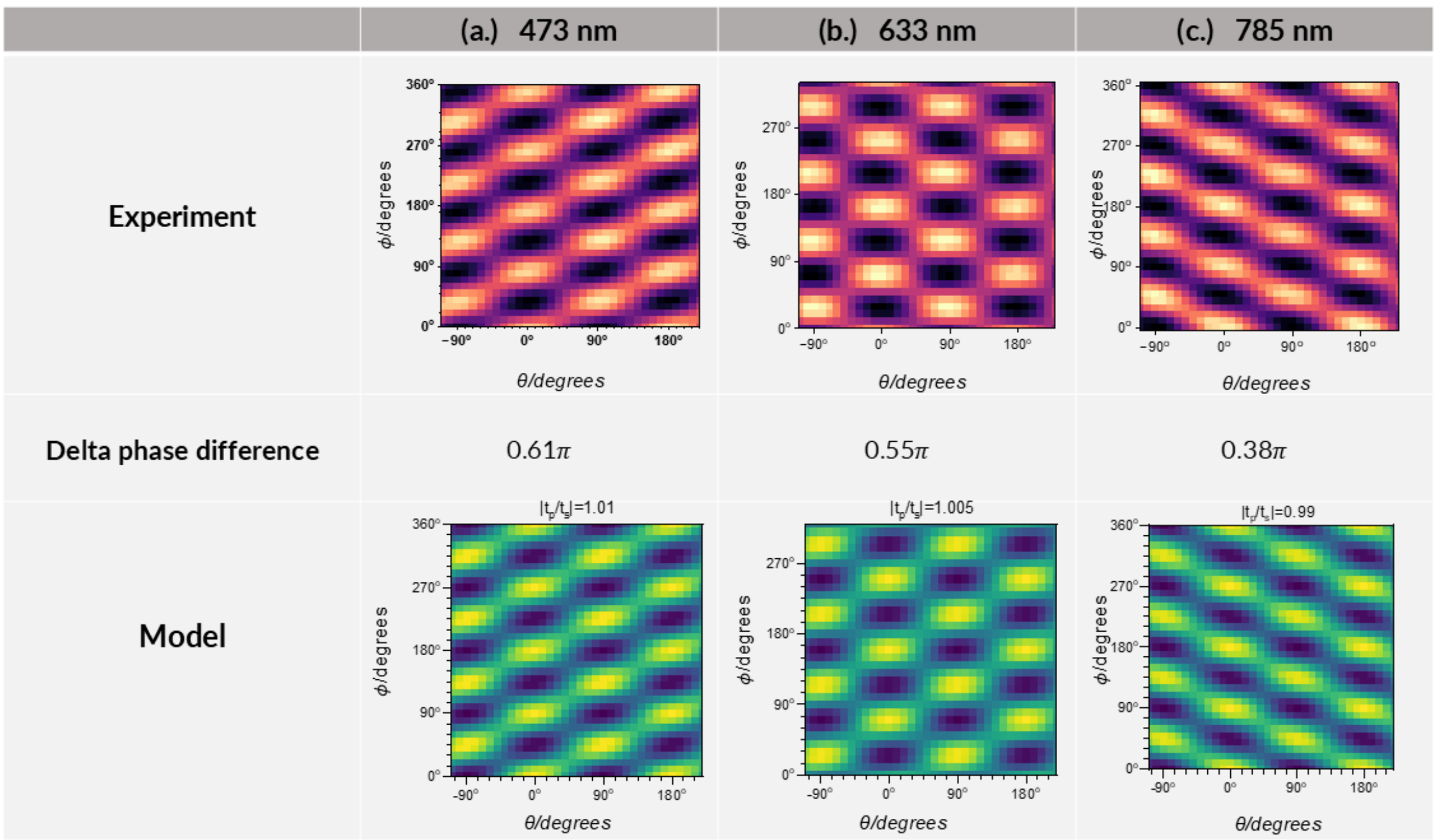

**Figure S12.** 2D polarization maps of the Si (100) $T_{2g}$ peak area at three excitation wavelengths show the experimental data and the corresponding data fitted with the model described in **Eq. (8)** in the main body of the paper. The differences in the delta phase across each excitation edge filter reveals how the polarizations are impacted differently.

For each excitation wavelength, we quantify how well the model reproduces the experimental results using the estimated variance shown in **Eq. S1**. The resulting values are: 0.019 (473 nm), 0.0045 (532 nm), 0.067 (633nm), and 0.0024 (785 nm). The model best reproduces the data for 785 nm.

## S12. ARRS anisotropic response 2D-Polarization maps of Si(100) using Volume Bragg Grating (VBG) Technology

In ARRS, it is a common practice to use VBG filters to study low-frequency phonons usually located below 100 cm$^{-1}$. To expand our insights and to explore the effect of these filters, we tested them using the anisotropic response of the Si (100), which previously unveiled the dichroic edge filter effect. Following the same protocol for collecting a 2D-PM, we used 532 nm as the excitation wavelength. **Figure S13a** shows the schematics of th**e** VGB filters and their placement on the setup. For the 2D-PMs, their range of for $\theta$ is from 0° to 180°, and for $\phi$ from 0° to 360°. **Figure S13b** is the experimental data of the Area of the Si(100) peak, and (c) the fitting of the experimental data using the ideal case equation (5) in the main paper, obtaining an estimated variance of 0.0007 (Eq. S3). These results highlight the absence of depolarization effects for Bragg filters.

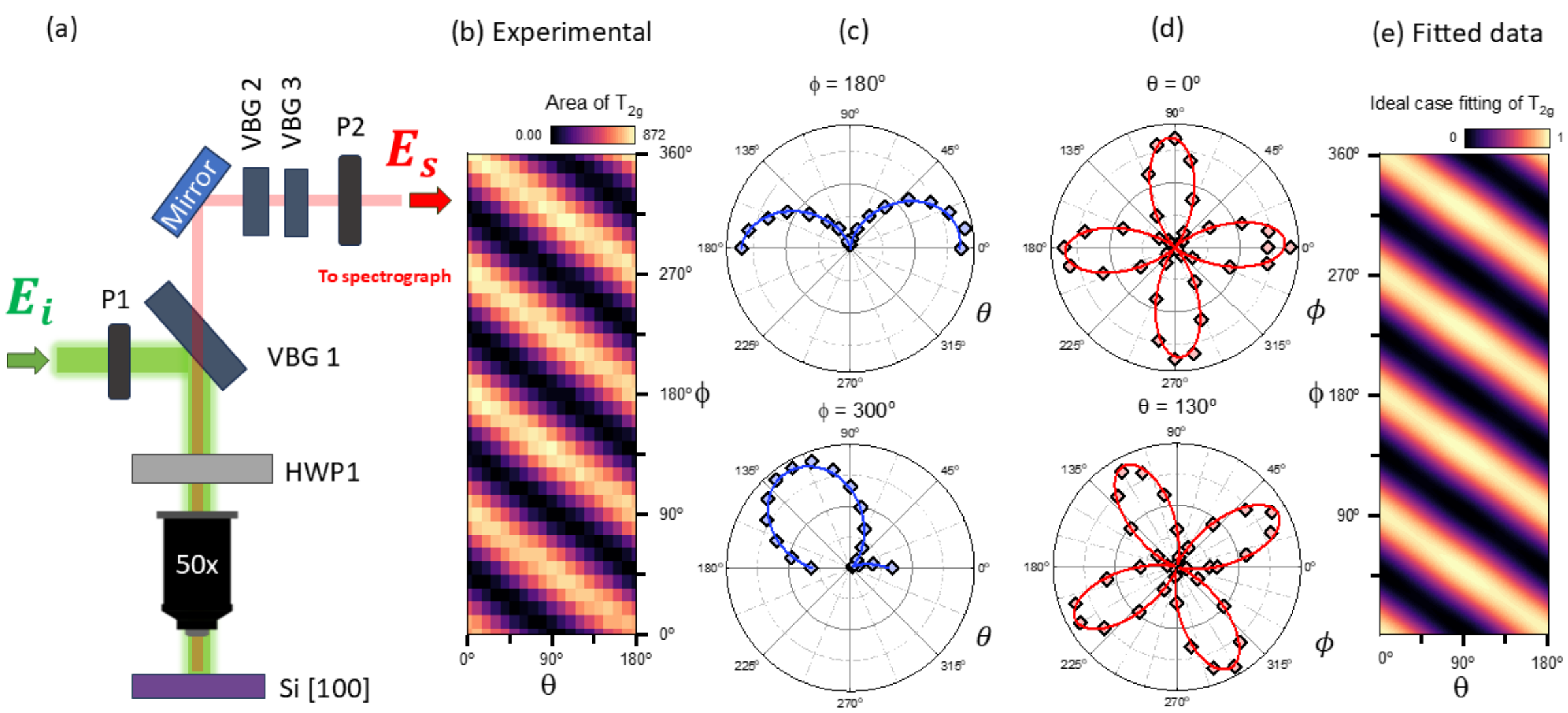

**Figure S13.**(a) Schematics of the VBG filter optics, where VBG1, VBG2, and VBG3 correspond to the different 3 filter matched to the excitation wavelength. Experimental (a) and fitted (b) 2D-PM of the $I_{T_{2g}}$ peak area as a function of the $\phi$ and $\theta$ in anisotropic Si(100): using Bragg filters with 532 nm excitation wavelength.(c) Polar plots at fixed $\phi$ = 180° and $\phi$ = 300° (diamonds) with an overlay of the theoretical prediction (blue solid line) as a function of θ rotation. (d) Polar plots at fixed θ = 0° and θ = 130° (diamonds) with an overlay of the theoretical prediction (red solid line) as a function of $\phi$ rotation.

Unlike the Si (100) polarization maps using the edge filter shown in the main body of the paper, the experimental $T_{2g}$ overlap with the theory at all angles, including angles such as $\phi = 300°$ and $\theta = 130°$, since the Bragg filters do not introduce any distortion in the polarization of the scattered light. The polarization distortion from the 532 nm excitation Bragg filters was collectively 0.10π with a variance of 0.007 (**Table S1**).

## S12. Characterization of the Super Achromatic Quarter-wave Plate (SA-QWP)

The response of the super achromatic quarter-wave plate (SA-QWP) was also characterized using a polarimeter for 473 nm, 532 nm, and 633 nm wavelengths. The ellipticity and light polarization ($\beta$) as a function of the SA-QWP rotation are shown in the **Figure S13**. We are interested in the QWP positions for which the light is fully circular (ellipticity ~ 45°), and its handiness is purely right or left. These two positions are separated by 90°. In the middle plot of **Figure S13** we can see that there is one position where the light is linear as the polarization through the QWP is the same as the fixed polarization set P1 in our system.

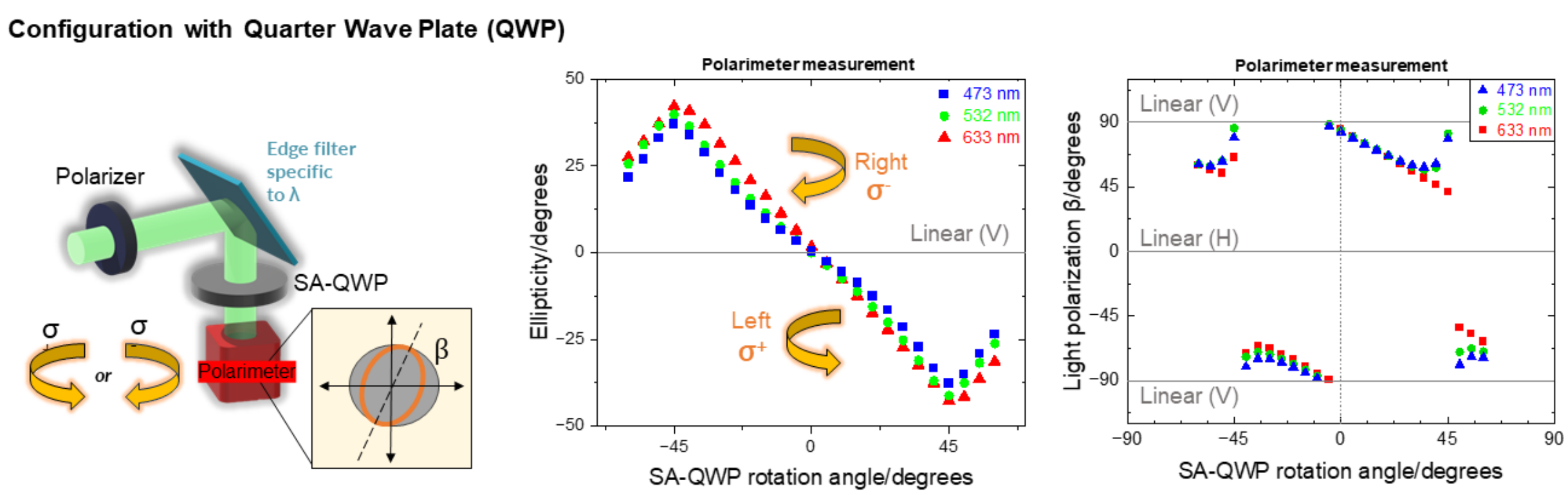

**Figure S14.** The circular polarization configuration showing the laser input and the polarimeter characterization of the super achromatic quarter-waveplate (SA-QWP). The zero ellipticity point corresponds to the linear polarization, whereas the inflection points at +45° and -45° correspond to the left-handed and right-handed circular polarization, respectively.

**S13. Calibration of the SA-QWP using the response of $A_{1g}$ mode of 2H-$MoS_2$**

Here we present the protocol of helicity-resolved Raman when there is no polarimeter available. In that case we use the Raman response of $A_{1g}$ and $E^1_{2g}/E_g$ modes from $MoS_2$ or $Al_2O_3$ under linear and circularly polarized light. First the response of $A_{1g}$ mode under linear polarization along the *x*-axis fixed by P1, is used to calibrate the HWP2, without any HWP or QWP placed before the objective. Following the step 6 in the flow chart presented in **Section 2.3 of the main body of the paper** to calibrate HWP2 and identify *parallel* ($\theta = 0°$), and *cross*($\theta = 90°$) configurations. Once the HWP2 is calibrated, the QWP is placed and calibrated using the helicity resolved $A_{1g}$ Raman response from $MoS_2$ or $Al_2O_3$.

The next step is set HWP2 to cross polarization configuration, where the $S/N$ of the $I_{A_{1g}}$ mode is minimized. Then, the QWP is rotated, and $I_{A_{1g}}$ will either peak or decrease as seen in **Figure S15**. The minimum response of the $A_{1g}$ mode (or maximum of the $E_{2g}$ mode in bulk $MoS_2$) defines the position where the QWP exhibits a "linear" polarization response; this position sets $\phi = 0^{\circ}$. It is important to note that if $MoS_2$ is used for calibration in the circular polarization configuration, the $I_{E_{2g}}$ will exhibit a $+\pi$ phase shift relative to the $I_{A_{1g}}$. Although the $S/N$ of the $E_{2g}$ mode is adequate for calibration, the $A_{1g}$ mode provides a significantly stronger S/N, making it ideal for fine adjustments smaller than 5 degrees. The maxima positions for $A_{1g}$ mode ($\pm 45°$ ) determine the position of the QWP, where the light passing through it is fully circularly polarized. However, without a polarimeter we cannot determine the handedness of the circular light. We used the QWP characterization by the polarimeter in the **Section S12** to label right ($\boldsymbol{\sigma_R}$) *R* and left ($\boldsymbol{\sigma_L}$) *L*. Regardless the distinguishability of the handedness of the circularly polarized light, without a polarimeter is possible to achieve co-rotating and cross-rotating configurations. the co-rotating *R-R* or *L-L*, response are measured with the HWP2 at $\theta = 90^0$ (and counter-rotating *L-R* or *R-L* with $0^0$). As we can see in **Figure 6b**, there is no difference in the Raman spectra between LR and RL, or RR and LL for the materials we use on our study

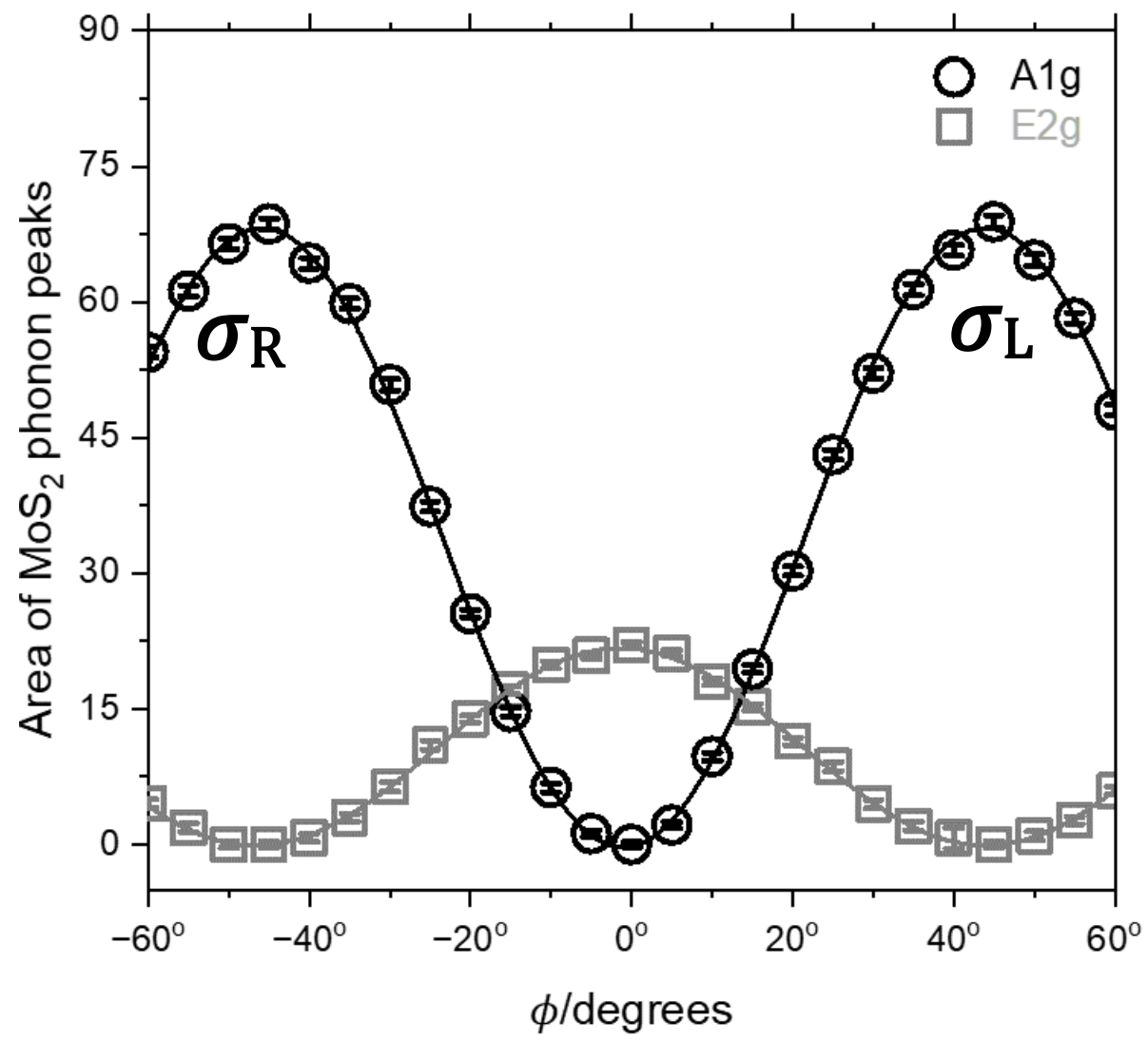

**Figure S15.** Plot of the fitted area of the $MoS_2$ phonons and their polarization response as a function of a super achromatic $\phi$-QWP angle in 5-degree increments (error bars show the standard error from Lorentzian peak area fitting). The minimum area of $A_{1g}$ ($\phi$ = 0°) corresponds to the "linear" polarization of the QWP at $\phi$ = 0°. From polarization handedness convention, the anticlockwise rotation of the QWP is the "left-handed" or $\sigma_L$ (where $\phi$ = +45°) and the clockwise rotation corresponds to the "right-handed" or $\sigma_R$ (where $\phi$ = -45°).